\documentclass[fleqn,usenatbib]{mnras}

\usepackage{graphicx}
\usepackage{txfonts}
\usepackage{multirow}
\usepackage{amssymb}
\usepackage{units}
\usepackage{float}
\usepackage{bm}   % Bold maths symbols, including upright Greek
\usepackage[dvipsnames]{xcolor}
\usepackage{color, colortbl}
\usepackage{xspace}
\usepackage{lscape}
\usepackage{url}
\usepackage{arydshln}

\usepackage[all]{hypcap}

\usepackage{enumitem,booktabs}
\usepackage{comment}
\usepackage{marginnote}
\usepackage[referable]{threeparttablex}
\renewlist{tablenotes}{enumerate}{1}

\newcommand{\FeKa}{Fe K\ensuremath{\alpha}\xspace}

\newcommand{\NH}{\ensuremath{N_{\mathrm{H}}}\xspace}

\newcommand{\xmm}{XMM{\it-Newton}\xspace}
\newcommand{\chandra}{{\it Chandra}\xspace}

\newcommand{\swift}{{\it Swift}\xspace}
\newcommand{\nicer}{{\it NICER}\xspace}
\newcommand{\rxte}{{\it RXTE}\xspace}

\newcommand{\xrism}{{XRISM}\xspace}
\newcommand{\athena}{{Athena}\xspace}
\newcommand{\arcus}{{ARCUS}\xspace}

\newcommand{\Hb}{H\ensuremath{\beta}\xspace}

\newcommand{\zw}{{$\rm I\:Zw\,1$}\xspace}

\newcommand{\nustar}{{\it NuSTAR}\xspace}

\newcommand{\spex}{{\textsc{Spex}}\xspace}

\newcommand{\ang}{{\AA}\xspace}

\newcommand{\ergflux}{{\ensuremath{\rm{erg\ cm}^{-2}\ \rm{s}^{-1}}}\xspace}

\DeclareRobustCommand{\ion}[2]{\textup{#1\,\textsc{\lowercase{#2}}}}
\newcommand{\civ}{\ion{C}{iv}\xspace}

\newcommand{\nv}{\ion{N}{v}\xspace}

\newcommand{\oi}{\ion{O}{i}\xspace}

\newcommand{\oiii}{\ion{O}{iii}\xspace}

\newcommand{\feii}{\ion{Fe}{ii}\xspace}

\newcommand{\pion}{\texttt{pion}\xspace}

\DeclareRobustCommand{\VAN}[3]{#2}
\let\VANthebibliography\thebibliography
\def\thebibliography{\DeclareRobustCommand{\VAN}[3]{##3}\VANthebibliography}

\title[The multi-epoch X-ray tale of I Zwicky 1 outflows]{The multi-epoch X-ray tale of I Zwicky 1 outflows}

\author[Daniele Rogantini et al.]{
D. Rogantini,$^{1,2}$\thanks{E-mail: danieler@mit.edu}
E. Costantini,$^{2,3}$
L. C. Gallo,$^{4}$
D. R. Wilkins,$^{5}$
W. N. Brandt,$^{6,7,8}$
and M. Mehdipour$^{9}$
\\
$^{1}$MIT Kavli Institute for Astrophysics and Space Research, Massachusetts Institute of Technology, Cambridge, MA 02139, USA\\
$^{2}$SRON Netherlands Institute for Space Research, Niels Bohrweg 4, 2333 CA Leiden, The Netherlands\\
$^{3}$Anton Pannekoek Astronomical Institute, University of Amsterdam, P.O. Box 94249, 1090 GE Amsterdam, the Netherlands\\
$^{4}$Department of Astronomy and Physics, Saint Mary’s University, Halifax, NS. B3H 3C3, Canada\\
$^{5}$Kavli Institute for Particle Astrophysics and Cosmology, Stanford University, 452 Lomita Mall, Stanford, CA 94305, USA\\
$^{6}$Department of Astronomy and Astrophysics, 525 Davey Lab, The Pennsylvania State University, University Park, PA 16802, USA\\
$^{7}$Institute for Gravitation and the Cosmos, The Pennsylvania State University, University Park, PA 16802, USA\\
$^{8}$Department of Physics, 104 Davey Lab, The Pennsylvania State University, University Park, PA 16802, USA\\
$^{9}$Space Telescope Science Institute, 3700 San Martin Drive, Baltimore, MD 21218, USA
}

\date{Last updated 2022 June; in original form 2022 June}

\pubyear{2022}

\begin{document}
\label{firstpage}
\pagerange{\pageref{firstpage}--\pageref{lastpage}}
\maketitle

% \abstract{}{}{}{}{} 
% 5 {} token are mandatory
 
\begin{abstract}
  {The narrow-line Seyfert 1 galaxy $\rm I\;Zwicky\,1$ shows a unique and complex system of ionised gas in outflow, which consists of an ultra-fast wind and a two-component warm absorber. In the last two decades \xmm monitored the source multiple times enabling the study of the long-term variability of the various outflows. Plasma in photoionisation equilibrium with the ionising source responds and varies accordingly to any change of the ionising luminosity. However, detailed modelling of the past RGS data has shown no correlation between the plasma ionisation state and the ionising continuum, revealing a complex long-term variability of the multi-phase warm absorber. Here, we present a new observation of $\rm I\;Zwicky\,1$ by \xmm taken in early 2020 characterised by a lower X-ray flux state. The soft X-ray spectrum from the RGS reveals the two components of the warm absorber with $\log \xi \sim -1.0$ and $\log \xi \sim 1.7$. Comparing our results with the previous observations, the ionisation state of the two absorbing gas components is continuously changing, following the same unpredictable behaviour. The new results strengthen the scenario in which the ionisation state of the warm absorber is driven by the density of the gas rather than the ionising luminosity. In particular, the presence of a radiation driven, inhomogeneous clumpy outflow may explain both the variability in ionisation throughout the years and the line-locked \nv system observed in the UV band. Finally, the EPIC-pn spectrum reveals an ultra-fast wind with an outflow velocity of $\sim 0.26c$ and ionisation parameter of $\log \xi \sim 3.8$.}
\end{abstract}

%
%-------------------------------------------------------------------

\section{Introduction}

A super-massive black hole with mass between $M_{\rm BH} \sim 10^6-10^{10}\ M_{\odot}$ and accreting material through an accretion disc is the core engine of an active galactic nucleus (AGN). Although the hearts of these powerful sources are the same, their phenomenology shows an extended variety based on several parameters such as line-of-sight angle and mass accretion rate \citep{Giustini19}. Around $\sim 50-65\%$ of nearby, bright, AGN show narrow absorption lines in their UV and X-ray spectra \citep[e.g.,][]{Costantini10,Kriss12,Tombesi13,Laha14}.
The plethora of absorption edges and lines imprinted by several ions in the soft X-ray band ($0.2-2$ keV) are known as a warm absorber and often show higher ionisation components compared to their UV counterparts \citep[e.g.,][]{Reynolds95,Costantini07,Mehdipour10,Behar17}. 

These absorption lines, predominately from astronomically abundant metals (e.g. C, N, O, Ne and Fe), are observed to be blueshifted with respect to the systemic redshift. This indicates that the gas is outflowing. The ionisation state of the absorber is usually expressed by the ionisation parameter, $\xi$, defined as $\xi = L_{\rm ion} /nr^2$ where $L_{\rm ion}$ is the ionising luminosity, $n$ the density of the gas and $r$ the distance from the ionising source. In many AGN, the warm absorber is a multi-component medium with a wide range in ionisation parameter ($\xi \sim 0.1-1000$~erg~cm~s$^{-1}$), column density ($\NH=10^{19}-10^{23.5}$) and outflow velocity ($v_{\rm out}\sim 100-3000\ {\rm km\ s^{-1}}$) and is very rarely detected as a single component \citep[e.g.,][]{Kaastra02,Kaspi02,Detmers11,Laha14,Silva16}. 

In spite of the extensive knowledge acquired through high-resolution X-ray spectroscopy with the gratings aboard \chandra and \xmm, there are still pending questions regarding warm absorbers in AGN. For example, their relation with other kinds of outflows observed in AGN, such as ultra-fast outflows \citep{Tombesi10,King15} and obscurers \citep{Kaastra14}, is still discussed among the scientific community. Furthermore, the spatial extent of the warm absorbers and their distance relative to the central black hole are difficult to determine. The spatial location of the outflows holds crucial information on the launching mechanism of warm absorbers and on the AGN feedback in the surrounding environment \citep[e.g.,][]{Crenshaw12,Fabian12,Veilleux20}.

An approach to estimate the distance of the absorber to the central source is monitoring the response of the gas to changes in the ionising continuum. The ionisation state of the gas responds to any change of the ionising luminosity on a timescale which yields information regarding the density of the gas \citep{Krolik95,Nicastro99} and its distance from the central source (through the definition of the ionisation parameter). The time that an ionised gas needs to reach the photoionisation equilibrium following a variation in the ionising luminosity is known as equilibrium timescale and it is inversely proportional to the density of the plasma. Low-density gases reaches their equilibrium slowly and they can show a lag between the evolution of the ionisation state and the illuminating radiation. Several attempts have been made to map the warm absorber distribution along the line of sight using long observations of both broad- \citep[Mrk 509,][]{Kaastra12} and narrow-line Seyfert~1 \citep[NGC 4051][]{Krongold07}. These complex studies indicate distances between ~0.1-500~pc. 

In a handful of AGN, outflows show a deviation from the expected behaviour in which their ionisation state responds with a possible delay to variations in ionising luminosity. For example, the ionisation state of the warm absorber gas detected in MR~2251-178 and Mrk~335 does not correlate with the X-ray luminosity \citep{Kaspi04,Longinotti13}. In MR~2251-178, the absorbing material seems to respond instantly to the ionisation variations only on short time-scales ($\sim$weeks). The odd long-term variability of the warm absorber properties can be explained with a changing absorber consisting of material that enter and disappears from the line of sight on timescales of several months. Mrk~335 shows, instead, a complex warm absorber system where only the high-ionisation component correlates with the X-ray flux variability, unlike the low-ionisation absorbers.

In this work we study the uncanny behaviour of the warm absorber detected in I Zwicky 1 (hereafter \zw). The source is the prototype of a narrow-line Type I Seyfert galaxy\footnote{Narrow-line Type I Seyfert galaxies are classified by their optical properties. These sources show (1) Balmer lines of H only slightly broader than the forbidden lines with a full-width-half-maximum less than $2000\; \rm km \; s^{-1}$  \citep{Goodrich89}; (2) strong \feii emission lines; and (3) [\oiii] 5007 \ang to \Hb ratio less than 3 \citep[][]{Osterbrock85}. In X-rays, NLS1 display a strong soft excess \citep[see for example][]{Boller96} and significant variability (see \cite{Gallo18} for a review of NLS1 X-ray properties).}. By virtue of its high optical nuclear luminosity and low redshift ($z=0.061169$) \zw is also known as the closest quasar \citep{Springbob05}. It is both a bright \citep[$L_{\rm X,\, (0.3-10\ keV)} \sim 10^{44}\ \rm erg\ s^{-1}$;][]{Gallo04} and variable X-ray source \citep{Wilkins17}. 

The soft X-ray spectrum of \zw reveals the presence of a two-component warm absorber which has been detected in all three previous \xmm  observations. The properties of these outflows and their long-term variability have already been studied by \cite{Costantini07b} and \cite{Silva18}. Both works noticed that the ionisation states of the low- and high-ionisation components do not correlate with the ionising luminosity. This behaviour is hard to explain and makes the warm absorber of \zw one of a kind. Based on their observations, \cite{Silva18} proposed a phenomenological model consisting of a clumpy outflow where the two gas components are potentially part of the same outflowing cloud, only differing in density and degree of exposure to the central source. 

The hard spectrum of \zw shows the presence of an ultra-fast outflow (UFO). Analysing in detail the iron K band of the archival \xmm observations, \cite{Reeves19} reported evidence for a fast ($\sim c$) wide-angle wind which is commonly seen in systems that accrete at close to the Eddington limit, such as \zw \citep{Poquet04}. The UFO presence is detected in three out of four epochs, with an ionisation parameter $\log \xi \sim 4.9$. In the 2005 observation, the absorption features are contaminated by a complex emission spectrum in the iron region. Adopting a radiative disc wind model, the fast outflow in \zw shows a mechanical power within $5\%-15\%$ of Eddington, while its momentum rate is of the order of unity \citep{Reeves19}. Comparing with the upper limits placed by the IRAM observation of the CO emission on the energetics of the molecular gas  \citep{Cicone14}, \cite{Reeves19} concluded that the scenario of a powerful, large-scale, energy conserving wind in \zw can be ruled out. They suggested the presence of a low efficiency mechanism in transferring the kinetic energy of the inner wind to the large-scale molecular component. 

In this work we present the analysis of the latest high-resolution X-ray spectrum taken during the \xmm observation of 2020. The recently-acquired data and its reduction are presented in Section \ref{sec:obs_data}. We introduce the spectral energy distribution (SED) of \zw in Section \ref{sec:sed}. The broadband modelling is essential to determine the ionisation state of the outflows described in Section \ref{sec:wa}. The results and their implications are discussed and summarised in Section \ref{sec:discussion} and \ref{sec:conclusion}, respectively. 

The spectral analysis and modelling presented here are performed using the \spex fitting package \citep{Kaastra96,Kaastra20} version 3.06.01. All the spectra shown in this work are background subtracted and displayed in the observed frame. We use both the $C$-statistic \citep{Cash79} and Bayesian data analysis \citep{Bayes63} for spectral fitting and provide errors at $1\sigma$ confidence level. In our calculation we adopted the cosmological parameters as default in \spex: $H_0=70\ \rm{km}\ \rm{s}^{-1}\ \rm{Mpc}^{-1}$, $\Omega_{\Lambda}=0.70$ and $\Omega_{m}=0.30$. We assume proto-solar abundances of \cite{Lodders10} throughout this paper. 

%--------------------------------------------------------------------
\section{Observations and data reduction}
\label{sec:obs_data}
%--------------------------------------------------------------------

\zw was observed by \xmm \citep{Jansen01} for two consecutive orbits (revolution 3680 and 3681) on January 12 and January 13, 2020 as part of the study of the X-ray spectrum and variability of the source \citep{Wilkins21}. The two observations with obsID 0851990101 and 0851990201 (hereafter, observation 101 and 201, respectively) have a total nominal exposure time of 145.1 ks (75.8 ks and 69.3 ks, respectively). The \xmm data were processed using the Science Analysis System (SAS, version 20.0) and CALDB v4.9.4. 

The Reflection Grating Spectrometer \citep[RGS,][]{denHerder01} was used to study the absorption-line-rich soft X-ray spectrum of \zw. The RGS data were gathered in the Spectroscopy HER+SES mode. The data were processed using the \texttt{rgsproc} pipeline task; the data and the background were extracted using default selection regions. Furthermore, we filtered out time intervals with background count rates $> 0.2$~count~s$^{-1}$ in CCD number 9. The background filtering led to a loss of about 14.6 ks. The total net exposure time is then 120.5~ks with an average source count rate of $\rm 0.244\pm0.002\ count\ s^{-1}$. The data were used in the spectral range between 7~\ang ($E\sim1.8$~keV) and 37 \ang ($E\sim0.34$~keV) and they were binned using the optimised binning routine \texttt{obin} \citep{Kaastra17} implemented in the \spex fitting package. For the time-averaged spectral analysis, we merged the two datasets and combined RGS1 and RGS2 in one single OGIP file \citep{Arnaud92} using the \texttt{rgscombine} task.

For the spectral energy distribution modelling we fitted spectra from \xmm EPIC-pn and OM. The EPIC-pn instrument \citep{Struder01} was operated in Small-Window mode with the Thin Filter. The data were processed using the \texttt{epproc} pipeline task. Periods of high-flaring background for EPIC-pn (exceeding 0.4~count~s$^{-1}$) were filtered out while applying the \texttt{\#XMMEA\_EP} filter. We extracted a single event (\texttt{PATTERN==0}), high-energy (\texttt{PI>10000 \&\& PI<12000}) light curve from the event file of the entire chip to identify intervals of high-flaring background. The last 2.5 ks of the first orbit (obsid 0851990101) has been filtered out due to the contamination by particle background. The \xmm EPIC-pn spectra were extracted from a circular region centred on the source with a radius of 30 arcsec. The background was extracted from a nearby source-free region of radius 30 arcsec on the same CCD as the source. The pileup was evaluated to be negligible using the task \texttt{eppatplot}. The single and double events were selected for the EPIC-pn (\texttt{PATTERN<=4}). Instrumental response matrices were generated for the spectrum using the \texttt{rmfgen} and \texttt{arfgen} tasks. We applied the energy dependent correction by \cite{Furst22} to align the EPIC-pn and \nustar spectral shape. In specific, we added the \texttt{XRT3\_XAREAEFF\_0014.CCF} to our calibration files and invoked it explicitly by running \texttt{arfgen} with the flag \texttt{applyabsfluxcorr=yes}. The fitted spectral range for EPIC-pn is 0.3–10 keV and for the analysis we grouped the data following the optimal binning algorithm presented by \cite{Kaastra16}.

The Optical Monitor \citep[OM,][]{Mason01} photometric filters were operated in the Science User Defined image/fast mode. In both observations, OM images were taken with the V, B, U, UVW1, and UVW2 filters, with an exposure time between 2.3 and 4.4 ks for each image. The OM images of \zw were processed with the \texttt{omichain} pipeline with the standard parameters. The filter count rates and errors were extracted from a region with an aperture of $12^{\prime\prime}$ and written into a single spectral file using the \texttt{om2pha} task.

\zw was also observed continuously by \nustar between 11-16 January 2020 (ObsID 60501030002) for a total net exposure time of 233~ks. We followed the standard procedure using \textsc{nustardas} v. 1.9.2 to reduce the observations. The event lists from each of the focal plane module (FPM) detectors were cleaned and calibrated using the \textsc{nupipeline} task. Following \cite{Wilkins21}, we extracted the source photons from a circular region, 30 arcsec in diameter, centred on the source. The smaller 30 arcsec region, suitable for sources that are faint above 10~keV, was selected over the larger 60 arcsec region to maximise the signal to background ratio in the observation of \zw. Source and background spectra together with their response matrices were extracted and generated using the \textsc{nuproducts} tool. In the spectral analysis we fitted simultaneously the $10-50$~keV energy band of the separate \nustar spectra obtained from the FPMA and FPMB detectors.

All the reduced files were converted from the OGIP FITS format to the \spex format using the auxiliary \texttt{trafo} program of the \spex package.

%--------------------------------------------------------------------
\section{Spectral energy distribution}
\label{sec:sed}
%--------------------------------------------------------------------

%-------------------------------------------------------------
%                            Spectral energy distribution 2020
%-------------------------------------------------------------

   \begin{figure*}
   \centering
   \includegraphics[width=.75\hsize]{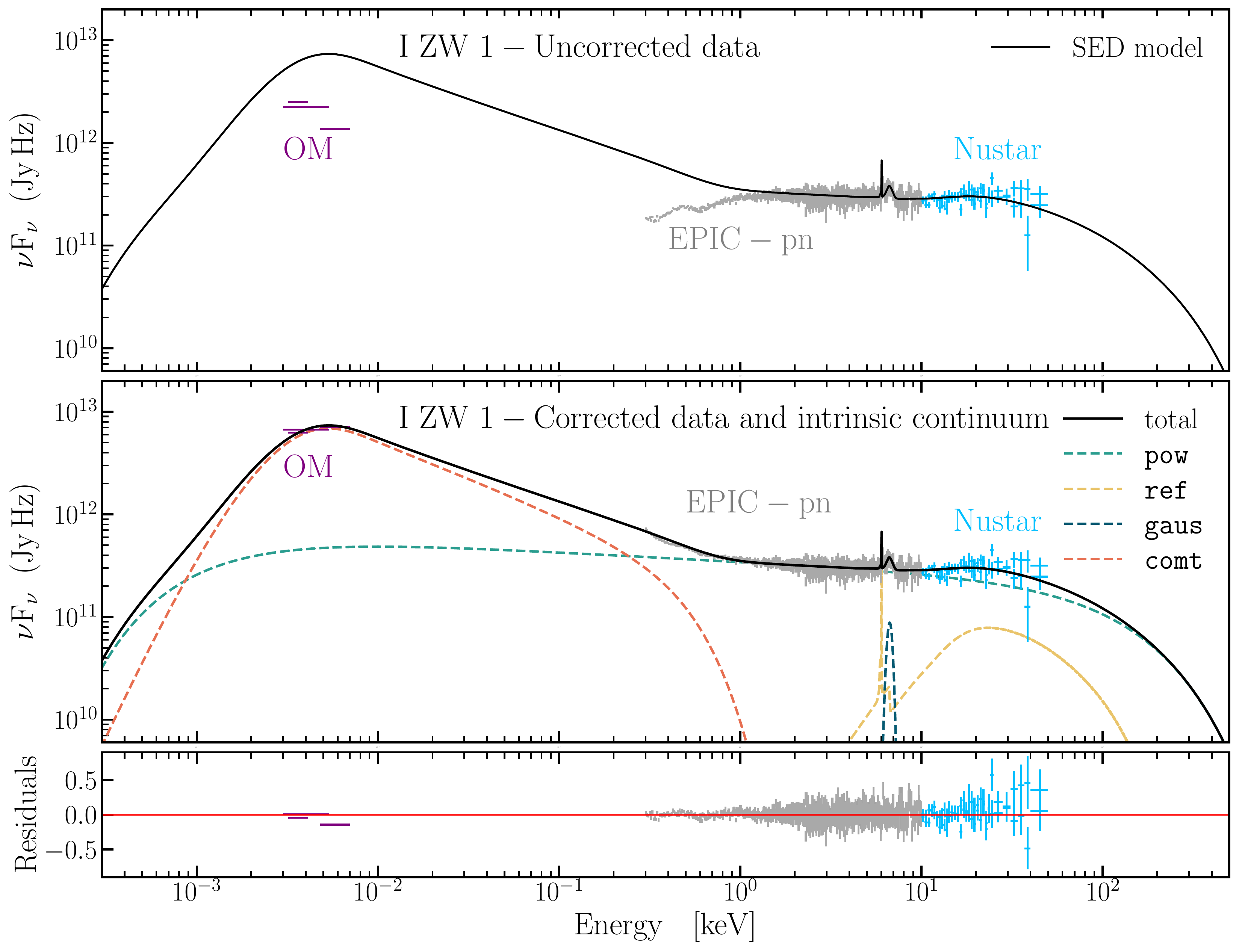}
      \caption{Average OM and EPIC-pn spectra of \zw from the \xmm observation taken in 2020 together with the FPMA and FPMB spectra from the simultaneous \nustar observation. The data are displayed before (\textit{upper panel}) and after (\textit{middle panel}) corrections for the intrinsic extinction and for the Galactic reddening in the UV (Section \ref{sec:sed_uv}) and in the X-rays (Section \ref{sec:sed_xray}). The data correction includes the effects of absorption by the warm absorber and the UFO. The different components used in our SED modelling are overplotted. The fit residuals, (observed-model)/model, are displayed in the \textit{bottom panel}. 
              }
         \label{fig:sed}
   \end{figure*}

In this section we present our modelling of the spectral components that form the observed SED of \zw. We jointly model the spectra of the two observations. We derive the intrinsic UV-X-ray continuum by modelling the UV reddening and X-ray absorption along our line of sight towards the nucleus of \zw. The final SED continuum model of the source is shown in Figure \ref{fig:sed}.

\subsection{UV continuum and reddening}
\label{sec:sed_uv}
To characterise the UV continuum emission we used the U, UVW1 and UVW2 filters of OM. The B and V filters are not considered in this analysis because of the uncertainties about the host-galaxy contribution at those wavelengths. In our first approach to model the UV emission radiated by the disc, we adopted a disk blackbody component ({\tt dbb} in \spex) with the temperature coupled to the seed temperature of the Comptonisation model \citep[\texttt{comt}, ][]{Titarchuk94} used to characterise the soft X-ray excess (Section \ref{sec:sed_xray}). The inclusion of the disc blackbody component does not improve the broad band fit and \texttt{comt} alone is sufficient to get a good fit of the UV/soft X-ray band. Therefore, we decided to used only the \texttt{comt} model to fit the UV continuum emission. A similar modelling has been used to model the optical/UV/soft X-ray emission from NGC~5548 and NGC~7469 \citep{Mehdipour15a,Mehdipour18}.

The line-of-sight extinction affecting \zw is the sum of the intrinsic absorption, in the rest frame of \zw, and a component due to our Galaxy, in the observed frame.
Assuming a \cite{Fitzpatrick99} reddening law with $R_V=3.1$, the foreground Milky Way reddening in our line of sight has a colour excess $E(B-V)=0.057$~mag \citep{Schlafly11}. We applied an \texttt{ebv} component in \spex to model this reddening, which incorporates the extinction curve of \cite{Cardelli89} including the update for near-UV given by \cite{odonnell94}.

\zw exhibits a steep UV/optical continuum which suggests a large degree of intrinsic reddening \citep{Laor97}. Using the flux measurements of the emission lines \oi $\lambda8446$ and \oi $\lambda 1304$, \cite{Rudy00} found an internal reddening of $E(B-V)=0.13$~mag. We therefore correct the UV emission adding a second \texttt{ebv} component with the colour excess parameter fixed at this value.

\subsection{X-ray continuum and absorption}
\label{sec:sed_xray}

%
%-------------------------------------------------------------
%                                                      SED FIT 
%-------------------------------------------------------------
%
\begin{table}
\caption{Best-value free parameters of the SED model \texttt{(pow+comt+refl+gaus)*red*pion*ebv$_{\tt HG}$*ebv$_{\tt MW}$*hot$_{\texttt{MW}}$}.}             
\label{tab:sed}
\centering          
\begin{tabular}{l c l }     % 3 columns 
\hline\hline       
\noalign{\vskip 0.75mm}

  Parameter &  Value & Units \\
\noalign{\vskip 0.75mm}
\hline
\noalign{\vskip 0.75mm}
  Norm/\texttt{pow}       & $1.95\pm0.02 $ &  $\rm 10^{52}\ ph\ s^{-1}\ keV^{-1}$ \\
  $\Gamma$/\texttt{pow}   & $2.09\pm0.01 $ &  $ $ \\
  \noalign{\vskip 0.75mm}
  Norm/\texttt{comt}      & $1.25\pm0.02 $ &  $\rm 10^{57}\ ph\ s^{-1}\ keV^{-1} $ \\
  $\rm t_1$/\texttt{comt} & $0.12\pm0.01 $ &  $\rm keV $ \\
  \noalign{\vskip 0.75mm}
  Scal/\texttt{refl}      & $0.43\pm0.08 $ &  $ $ \\
  \noalign{\vskip 0.75mm}
  Norm/\texttt{gaus}      & $6\pm1 $ &  $\rm 10^{49}\ ph\ s^{-1} $ \\
  $E_0$/\texttt{gaus}     & $7.07\pm0.06 $ &  $\rm keV $ \\
  FWHM/\texttt{gaus}      & $0.6\pm0.1 $ &  $\rm keV $ \\
    \noalign{\vskip 0.75mm}
\hline
  \noalign{\vskip 0.75mm}
  $F_{0.2-2\ \textrm{keV}}$ & $4.76\pm0.01 $ &  $10^{-12}\ \ergflux $ \\
  $F_{2-10\ \textrm{keV}}$  & $4.73\pm0.02 $ &  $10^{-12}\ \ergflux $ \\
    \noalign{\vskip 0.75mm}
\hline 
  \noalign{\vskip 0.75mm}
  $C$stat/dof  & $993/902$ &   \\
  \noalign{\vskip 0.75mm}
\hline                 
\end{tabular}
%\tablefoot{$^{\dag}$ The dispersion velocity of the third component has been fixed to 80 km s$^{-1}$.}
\end{table}

To fit the time-averaged X-ray spectrum of \zw we started with a power-law continuum (\texttt{pow} component in \spex). The model mimics the Compton-up scattering of the disc photons in a optically-thin, hot, corona. The high-energy exponential cut-off is fixed to 140 keV \citep{Wilkins22}. A similar exponential cut-off has been applied to the power-law continuum in the low-energy band ($E_{\rm cut}=13.6$~eV) in order to prevent it from exceeding the maximum energy of the seed disc photons. {\bf We coupled the energy of the low cut-off with the temperature, $T_{\rm seed}$ of the seed photon of the Comptonisation model.} The photon index $\Gamma$ of the intrinsic power-law is $2.09\pm0.01$. 

The soft X-ray continuum of \zw shows the presence of an excess above the power-law \citep[][]{Gallo04,Gliozzi20}. Previous works fitted this excess using a broken power-law with an energy break at 2~keV \citep{Costantini07,Silva18}. The full modelling of the X-ray continuum of the 2020 observation of \zw has been presented in \cite{Wilkins22}. There the data showed a preference for a reflection model from an accretion disc with a radial ionisation component which can well describe at the same time the soft X-ray excess, the Fe K$\alpha$ line and the Compton hump. Here, we aim at a simpler best fit, that models well the data while keeping a fast computational time of the photoionised code (see Section \ref{sec:wa}. We modelled the soft excess adopting the Comptonization model which fits well the emission fluxes in the U, UVW1 and UVW2 filters of OM, as shown before. In this interpretation of the soft excess, the seed disc photons are up-scattered in a warm, optically thick, corona to produce the extreme ultraviolet emission and the soft X-ray excess as its high-energy tail \citep[e.g.,][]{Done12,Petrucci18,Kubota18}. In their modelling, \cite{Wilkins22} noticed that the soft-excess can be equally well described by the emission from a warm corona that extends over the inner region of the accretion disc.
%this optically thick corona Comptonizes the thermal photons emitted from the accretion disc, but instead of producing a power law spectrum extending to high energies, the warm corona produces an excess of soft Xray emission, peaking below 1 keV
%We modelled the soft excess adopting a Comptonisation model (\texttt{comt} in \spex), which has already been used to model the soft X-ray continuum in similar Seyfert-1 AGN, such as Mrk~509 \citep{Mehdipour11} and NGC~3227 \citep{Mehdipour21}. As already shown in the previous section, this Comptonisation model fits well the emission fluxes in the U, UVW1 and UVW2 filters of OM. In this interpretation of the soft excess, the seed disc photons are up-scattered in a warm, optically thick, corona to produce the extreme ultraviolet emission and the soft X-ray excess as its high-energy tail \citep[][and references therein]{Kubota18}. However, in our SED modelling, we are only interested in a phenomenologically representation of the soft excess. For a detailed analysis and full discussion, we refer to \cite{Wilkins22}.

The parameters of the \texttt{comt} model are its normalisation, seed photon temperature ($T_{\rm seed}$), electron temperature ($T_{e}$), and optical depth ($\tau$) of the upscattering plasma. In order to avoid degeneracy between the plasma parameters of the X-ray soft excess and the UV emission, we fixed $T_{\rm seed}$ to a fiducial value of 1~eV and the optical depth to a value of 20. This limits the number of free parameters while still providing a good fit.

Subsequently, we added an X-ray reflection component (\texttt{refl} in \spex) to take into account the reprocessing of the incident X-ray continuum, which is evident by the presence of the \FeKa line at 6.4 keV \citep{Leighly99,Reeves00} and the Comptonisation hump. The \texttt{refl} component computes the Compton-reflected continuum \citep{Magdziarz95} and the \FeKa line \citep{Zycki99}. Similarly to the observed primary power-law, the high-energy exponential cut-off of the incident power-law component has been set to 140~keV. We adopted solar metal abundances and we fixed the ionisation parameter of \texttt{refl} to the lower limit of zero in order to produce a cold reflection component. Furthermore, we coupled the photon index $\Gamma$ of the incident power-law with the one of the primary power-law and we fitted the reflection scale factor ($s$). 

Previous observations of \zw have shown the presence of a broad ionised emission line in the iron K band centred at $E=7.0$~keV \citep{Gallo07b,Reeves19}. Thus, we added a Gaussian profile (\texttt{gaus} in \spex) to fit the emission from the partially ionised component. The normalisation, the line energy ($E_0$) and the full width at half maximum of the Gaussian line have been computed. 

%-------------------------------------------------------------
%                                          Residuals
%-------------------------------------------------------------
   \begin{figure*}
   \centering
   \includegraphics[width=\hsize]{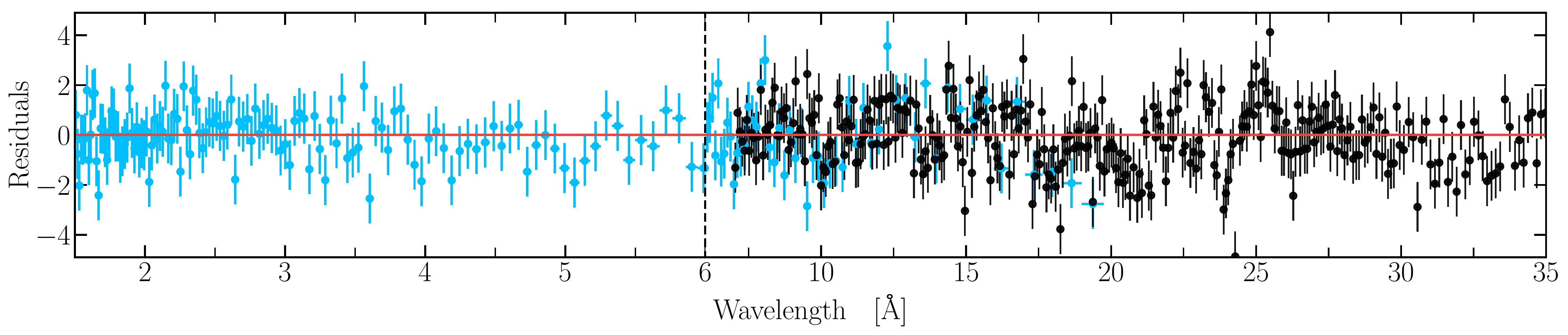}
   \vspace{-0.5cm}
      \caption{Residuals of the RGS and EPIC-pn spectra (in black and blue, respectively) obtained fitting the 2020 \xmm observation without any outflow component. We modified the $x$-scale in order to show clearly the residuals in both instruments. The dashed vertical line indicates where the scale changes. The residuals are defined as (data-model)/error.
              }
         \label{fig:res}
   \end{figure*}

In our SED modelling, we take into account the X-ray continuum and line absorption by the foreground diffuse interstellar medium in the Milky Way. We used the \texttt{hot} component in \spex \citep{dePlaa04,Steenbrugge05}, which calculates the transmission of a plasma in collisional ionisation equilibrium at a given temperature. To mimic the absorption by the intervening cold interstellar medium we set the temperature to its minimum, $kT=0.008$~eV ($T\sim 100\; \rm K$). The hydrogen column density was fixed to $N_{\rm H}^{\rm MW} = 6.01\times10^{20}\ \textrm{cm}^{-2}$ \citep{Elvis89,Willingale13}, which includes both the atomic and molecular hydrogen component. 

Absorption by ionised AGN outflows significantly affects the X-ray spectral shape of \zw \citep[][]{Silva18}. In our spectral modelling we take into account absorption by persistent outflows adopting the \texttt{pion} component in \spex \citep[][]{Mehdipour16}. This model calculates the absorption spectrum from physical absorber parameters on the fly and therefore does not require any model grid preparation before data fitting. The warm-absorber models used here to fit the EPIC-pn data are based on a preliminary analysis of the RGS dataset. We decided to neglect the high-ionisation warm absorber component since it has a negligible impact on the spectral shape of the SED (see Section \ref{sec:wa}). The column density and the ionisation parameter $\xi$ \citep{Krolik81} of the low-ionisation warm absorber and UFO were freed in order to improve the fit of the EPIC-pn data. 

The best-fit parameters of the power-law component (\texttt{pow}), warm Comptonisation component (\texttt{comt}), the X-ray reflection component (\texttt{refl}), the Gaussian component (\texttt{gaus}) for the blend of H- and He-like \FeKa lines are provided in Table \ref{tab:sed}. The best-fit to the data is shown in Figure \ref{fig:sed}.

%--------------------------------------------------------------------
\section{Outflows}
\label{sec:wa}
%--------------------------------------------------------------------

%
%-------------------------------------------------------------
%                                               Warm Absorbers 
%-------------------------------------------------------------
%
\begin{table*}
\caption{\bf Best-fit parameters of the warm absorber components of \zw in the 2020 time-averaged spectrum inferred with a standard $C$-statistic analysis (on the left) and Bayesian framework (on the right).}             
\label{tab:wa}
\centering
\renewcommand{\arraystretch}{1.25}  
\begin{tabular}{c c c c c c | c c c c}     % 8 columns 
\hline\hline       

 & \NH & $\log \xi$ & $v_{\rm turb}$ & $v_{\rm out}$ & $C$-stat/dof  &\NH & $\log \xi$ & $v_{\rm out}$ & $\log Z$ \\
 &\scriptsize $10^{20}\ \rm cm^{-2}$ &\scriptsize  $\rm erg\ cm\ s^{-1}$ &\scriptsize  $\rm km\ s^{-1}$ & \scriptsize  $\rm km\ s^{-1}$ & &\scriptsize $10^{20}\ \rm cm^{-2}$ &\scriptsize  $\rm erg\ cm\ s^{-1}$ &\scriptsize  $\rm km\ s^{-1}$ & \\
\hline
\multicolumn{9}{c}{Single warm absorber component} \\
\hline
 LIC & $7.7_{-0.5}^{+0.8}$ & $-1.0\pm0.1$ & $ 110_{-25}^{+33}$ & $-1750\pm100$ & $3218/2433$ & $9.4\pm0.6$ & $-1.2\pm0.1$ & $-1750\pm100$ & $-6.6$\\
\hline
\multicolumn{9}{c}{Single warm absorber component + ultra-fast outflow} \\
\hline
 LIC & $9.0\pm0.7$ & $-1.0\pm0.1$ & $ 90_{-20}^{+40}$ & $-1750\pm100$ & \multirow{2}{*}{$3173/2430$} & $10.0\pm0.5$ & $-1.2\pm0.1$ & $-1750\pm100$ & \multirow{2}{*}{$0.0$}\\
 UFO & $200_{-100}^{+300}$ & $3.8\pm0.1$ & $100^{\dag}$ & $-77100\pm400$ & & $200\pm50$ & $3.8\pm0.1$ & $-77000\pm3000$ & \\
\hline
\multicolumn{9}{c}{Two warm absorber components + ultra-fast outflow} \\
\hline
LIC & $9\pm1$ & $-1.0\pm0.1$ & $110\pm30$ & $-1750\pm100$ & \multirow{3}{*}{$3162/2427$} & $9.4\pm0.5$ & $-1.2\pm0.1$ & $-1750\pm100$ &\multirow{3}{*}{$-0.3$} \\
HIC & $1.0\pm0.6$ & $1.7\pm0.2$ & $100^{\dag}$ & $-2150_{-250}^{+200}$ & & $0.9\pm0.3$ & $1.8\pm0.3$ & $-2150\pm350$ & \\
UFO & $200_{-100}^{+300}$ & $3.80_{-0.04}^{+0.11}$ & $100^{\dag}$ & $-77100\pm400$ & & $200\pm50$ & $3.8\pm0.1$ & $-77000\pm3000$ & \\
\hline
\end{tabular}
\renewcommand{\arraystretch}{1.}
\flushleft{\footnotesize{$^{\dag}$The dispersion velocity has been fixed to the default value of 100 km s$^{-1}$.}}
\end{table*}

%-------------------------------------------------------------
%                                               Warm absorbers
%-------------------------------------------------------------
   \begin{figure*}
   \centering
   \includegraphics[width=.75\hsize]{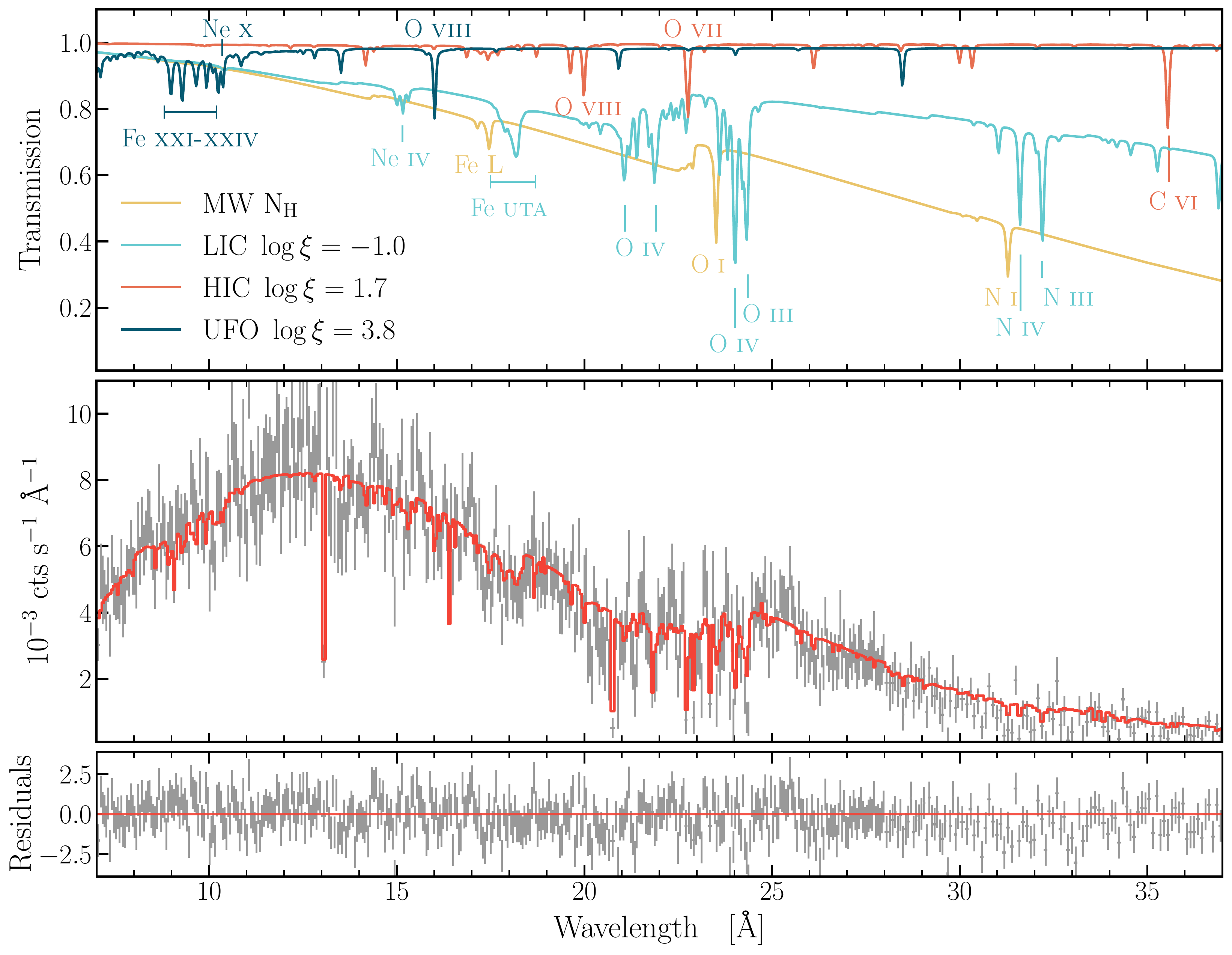}
      \caption{Combined RGS observations of \zw taken in 2020. \textit{Upper panel:} transmission spectra of the detected warm absorbers (light-blue and red line), ultra-fast outflow (dark blue) and foreground neutral Galactic interstellar medium (yellow line). We labelled the strongest absorption features of the models. \textit{Middle panel:} best fit of the RGS spectra of \zw. The spectrum has been binned for clarity of presentation. \textit{Lower panel:} residuals of the best fit, defined as $\rm (data-model)/error$. 
              }
         \label{fig:wa}
   \end{figure*}

%-------------------------------------------------------------
%                                               Warm absorbers
%-------------------------------------------------------------
   \begin{figure*}
   \centering
   \includegraphics[width=.75\hsize]{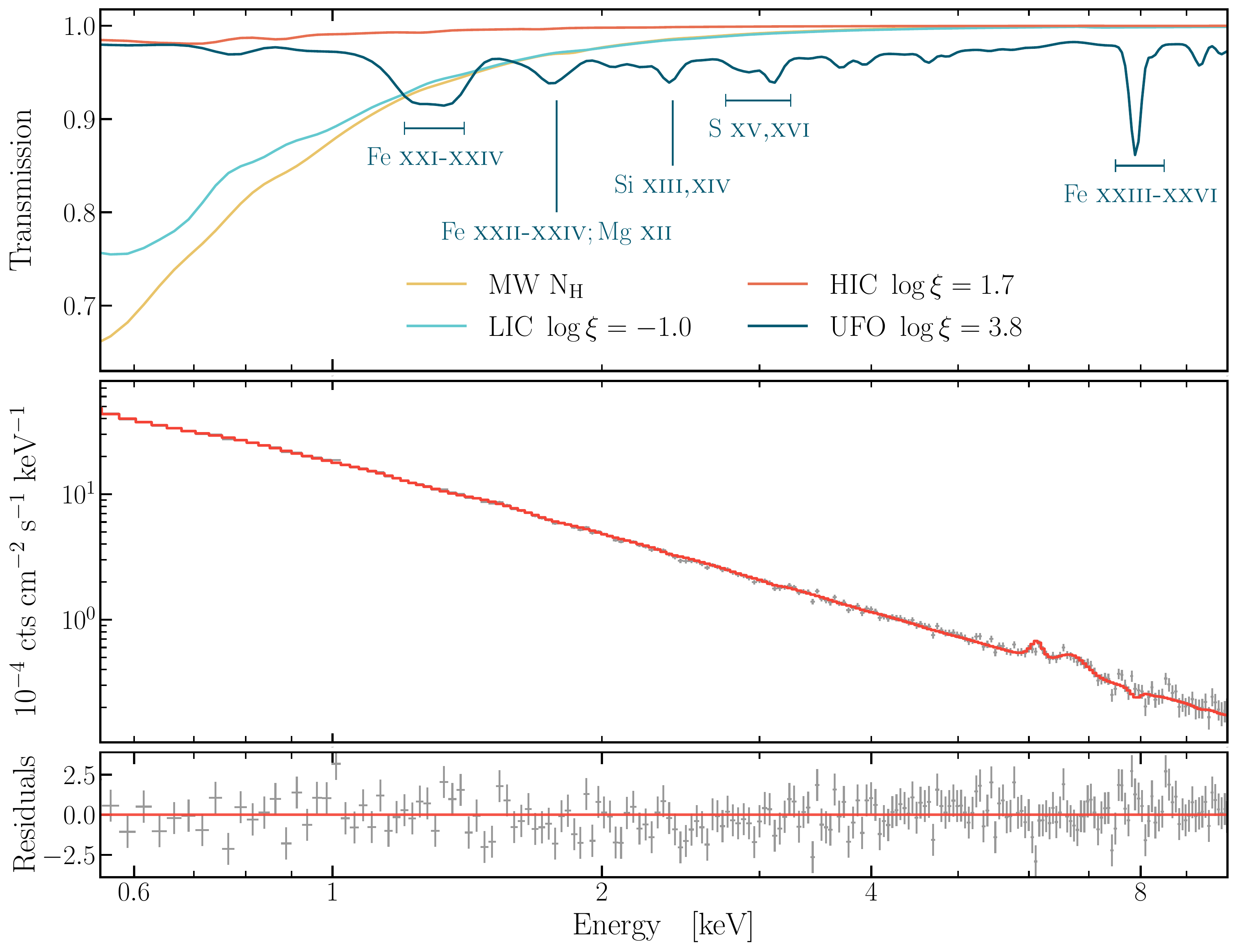}
      \caption{Combined PN observations of \zw taken in 2020. \textit{Upper panel:} transmission spectra of the detected warm absorbers (light-blue and red line) and ultra-fast outflow (dark blue). The foreground Galactic neutral absorption is also shown (yellow line). We labelled the strongest absorption features of the models. \textit{Middle panel:} best fit of the PN spectra of \zw. The spectrum has been binned using optimised binning \citep{Kaastra17}. \textit{Lower panel:} residuals of the best fit, defined as $\rm (data-model)/error$. 
              }
         \label{fig:wa_pn}
   \end{figure*}

%-------------------------------------------------------------
%                                          EPIC-pn light curve
%-------------------------------------------------------------
   \begin{figure*}
   \centering
   \includegraphics[width=.75\hsize]{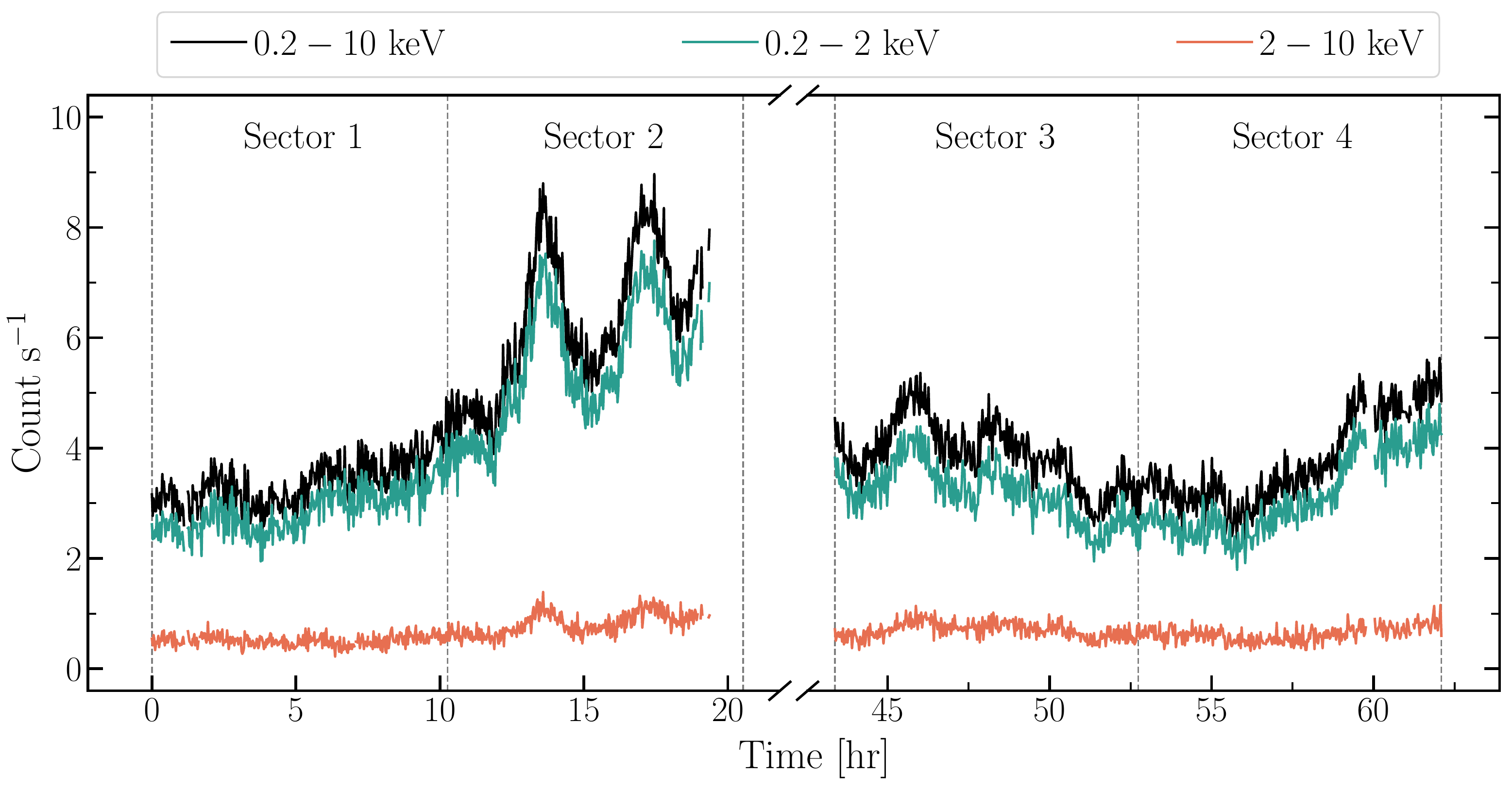}
      \caption{EPIC-pn light curve of the \zw observations taken 2020. We overplot the light curve extracted in different energy ranges. $\rm{Time} = 0$ corresponds to the starting time of the first observation (MJD = 58860.94956941). The dashed vertical lines delimit the extraction sectors used for the time-resolved spectral analysis.
              }
         \label{fig:pn_lc}
   \end{figure*}

We present here the AGN wind and its long-term variability through photoionisation modelling and \xmm RGS spectroscopy. For the spectral fitting of the absorption lines detected in the soft X-ray band we used multiple \texttt{pion} components with elemental abundances fixed to the proto-solar values.

%--------------------------------------------------------------------
\subsection{Time-averaged spectrum}
\label{sec:tave_wa}
To characterise the nuclear ionised outflows of \zw, we adopted the photoionisation continuum presented in Section \ref{sec:sed_xray} and we analysed in detail the time-averaged RGS and EPIC-pn spectra of \zw. In Figure \ref{fig:res}, we show the residuals obtained fitting the data only with a galactic absorbed continuum model. In the energy band covered, there are several visible absorption features which are the typical signatures of intervening ionised absorbers. Firstly, we added a \texttt{pion} component to account for the photoionisation features impressed by the warm absorber. Given an SED the model computes both the photoionisation solution and the transmitted spectrum of a plasma in photoionisation equilibrium. The derived shape of the SED can indeed significantly influence the structure and the thermal stability of the AGN outflows \citep[see e.g.,][]{Chakravorty12}. The parameters of interest of the \texttt{pion} component are the hydrogen column density (\NH), ionisation parameter (usually expressed in logarithmic scale, $\log \xi$), turbulence velocity ($v_{\rm turb}$), and outflow velocity ($v_{\rm out}$) of the ionised outflow.

We observed a warm absorber component with a ionisation parameter $\log \xi = -1.0 \pm 0.1$ and column density $\NH = (9\pm1)\times10^{20}\rm \ cm^{-2}$. Both outflow velocity ($v_{\rm {out}} = -1750 \pm 100\rm \ km\ s^{-1}$) and turbulence velocity ($v_{\rm {turb}} = 110 \pm 30\rm \ km\ s^{-1}$) of this low-ionisation component (LIC hereafter) are comparable with the lower ionisation outflow observed in earlier datasets \citep{Costantini07,Silva18}. To investigate the multi-phase nature of the warm absorber we added another \texttt{pion} component to our spectral model looking for the higher ionisation warm absorber detected in earlier observations \citep[e.g.][]{Silva18}. We found an outflowing plasma with an ionisation parameter of $\log \xi = 1.7\pm0.2$, column density $\NH = (1.0\pm0.6)\times10^{20}\rm \ cm^{-2}$ and outflow velocity $v_{\rm out} = -2150^{+200}_{-250}\rm \ km\ s^{-1}$. However, the statistical significance of the high-ionisation component (HIC, hereafter) is low; the $C$ improves by $\Delta C\textrm{stat}= 11$ for 3 free parameters.

We also tested the presence of an ultra-fast outflow component. The time-averaged spectral analysis of the 2020 \xmm observation reveals the presence of a UFO with hydrogen column density $\NH = (2_{-1}^{+3})\times10^{22}\rm \ cm^{-2}$, outflow velocity $v_{\rm {out}} = -77100 \pm 400\rm \ km\ s^{-1}$ ($\sim0.26 c $) and ionisation parameter $\log \xi = 3.80_{-0.04}^{+0.11}$. A UFO-like feature was already detected by \citep{Wilkins22}. For this fast wind, the turbulence velocity was set to its default value of 100~km/s. Adding this component to the broad model the statistic improves by $\Delta C{\rm stat} = 45$ for 3 free parameters. We did not detect any strong absorption feature in the iron K band. The transmitted spectrum of this fast outflow mainly improve the fit at lower energies, between 1 and 4 keV where multiple transitions by Si, S and Fe are located.

We found a lower ionisation parameter with respect to the previous work by \cite{Reeves19} which obtained a $\log \xi = 4.91_{-0.13}^{+0.37}$. This large discrepancy cannot be explained by only considering the discrepancy between different photoionisation models and SED modelling. It might suggest instead that the ionisation state of the UFO varies over year-long timescales contrary to what has been observed in the previous \xmm epochs \citep{Reeves19}.

When defining the relation between the different components inside \spex, we ordered the ionised absorber components following the criterion that the higher ionisation outflows are located closer to the central source shielding the lower ionisation ones from the ionising luminosity. Thus, each outflow component (UFO and warm absorbers) sees a different ionising continuum. All the best-fit parameter values of our modelling are listed in Table \ref{tab:wa}. We sorted the \pion component by their significance. The best fit of the RGS and the EPIC-pn spectra are shown separately for clarity in Figure \ref{fig:wa} and \ref{fig:wa_pn}, respectively. For each ionised absorber, we also display the single transmittance spectral model which quantifies the respective contribution to absorption.

In parallel to the the X-ray spectral analysis based on the $C$-statistic, we characterised the AGN ionised outflows with Bayesian parameter inference. The Bayesian approach has the advantage of exploring the entire parameter space identifying sub-volumes which constitute the bulk of the probability. We used the MultiNest algorithm \citep[v3.10][]{Feroz09,Feroz13} and we adapted the \texttt{PyMultiNest} and Bayesian X-ray Analysis, \texttt{BXA}, packages \citep{Buchner14} to the \spex fitting code \citep[][]{Rogantini21}. 

To test the significance of each warm absorber component we run the analysis twice starting with a model with one single \pion component and then adding a second component. In both models we included an additional photoionisation model to characterise the ultrafast outflow. In order to minimize the number of free parameters and to speed up the analysis we fixed the turbulence velocity to its default value ($100\ \rm km \ s^{-1}$). We assumed uninformative priors for the column density, ionisation parameter and outflow velocity of each \texttt{pion} model. The corner plot of the final parameter distributions is displayed in Figure \ref{fig:bay} for the model with three \pion components. 

In addition, the Bayesian framework provides a robust model comparison based on the Bayesian evidence, $Z$, which represents the posterior probability of the model given the data \citep[see][and references therein for a detailed explanation]{Buchner14}. This method does not make any assumptions about the parameter space or the data. We computed and compared the evidence of the two models (with and without the HIC) adopting the scale of \cite{Jeffreys61} which excludes models with a Bayes factor of 30 (a difference of 1.5 in $\log Z$). Both models show a comparable significance and there is not any strong evidence against the second warm absorber component. We cannot, therefore, exclude the presence of the HIC. The parameter values inferred with the Bayesian approach are listed on the right side of the Table \ref{tab:wa} together with the Bayesian evidence of each model.

\subsection{Time resolved spectroscopy}
\label{sec:tres_wa}
%-------------------------------------------------------------
%                                   Time resolved spectroscopy 
%-------------------------------------------------------------
   \begin{figure}
   \centering
   \includegraphics[width=\hsize]{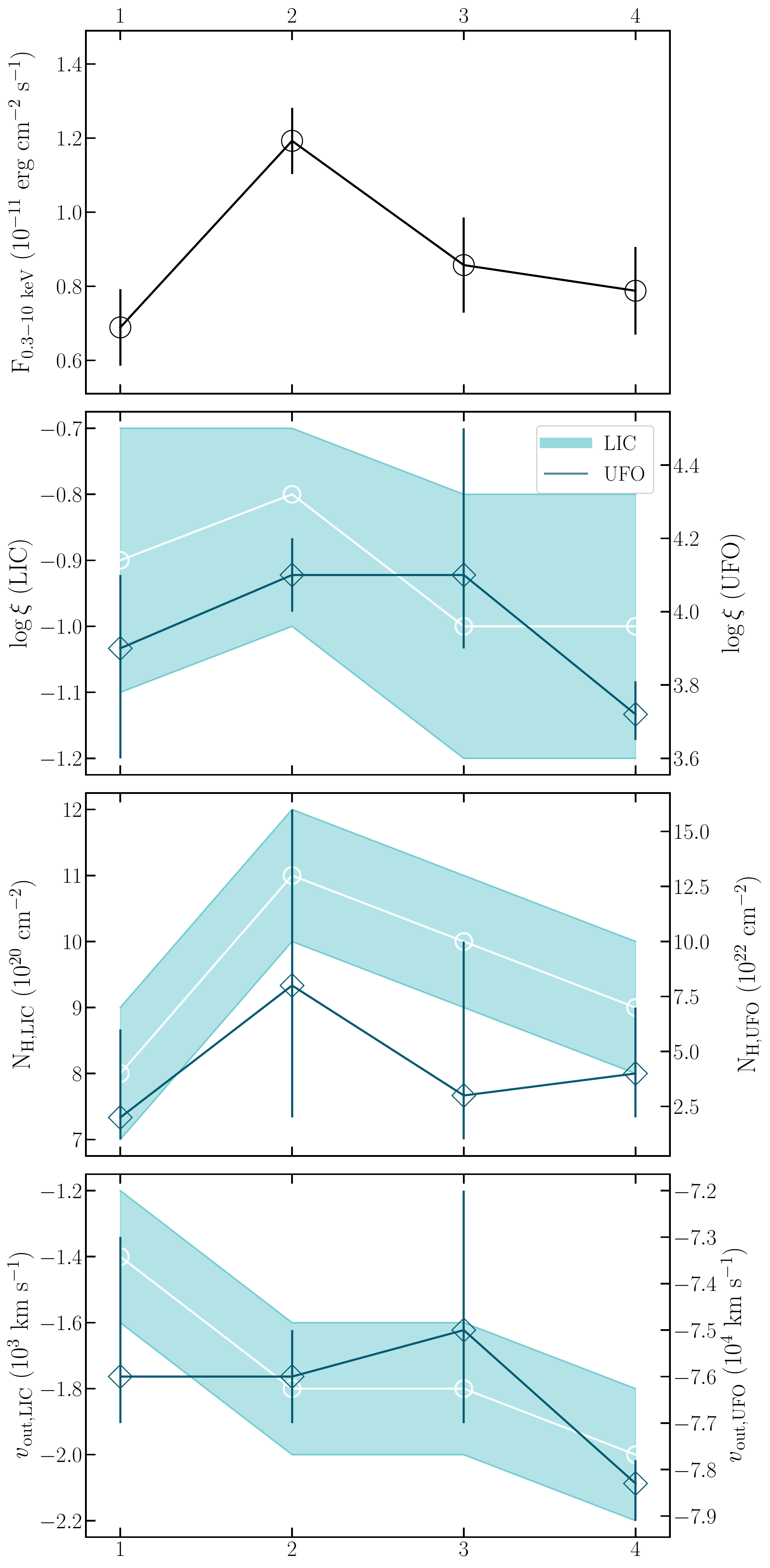}
      \caption{Time-resolved spectroscopy analysis. In the {\it top} panel we show the $0.3-10$~keV flux for each sector. In the second, third and fourth panels we plot respectively the evolution of the ionisation parameter, column density and outflow velocity for both the low-ionised warm absorber (shaded light blue, left $y$-axis) and the ultra-fast outflow (dark blue, right $y$-axis.) 
              }
         \label{fig:trs}
   \end{figure}

%
%-------------------------------------------------------------
%                                   Time-resolved spectroscopy
%-------------------------------------------------------------
%
\begin{table*}
\caption{Time-resolved spectroscopy: best-parameters of the outflows.}             
\label{tab:trs}
\small
\centering
\renewcommand{\arraystretch}{1.25}  
\begin{tabular}{c c c c c @{\hspace{0.1cm}}c c c c }     % 8 columns 
\hline\hline       
 & \multicolumn{3}{c}{LIC} & & \multicolumn{3}{c}{UFO} & \\ \cline{2-4}  \cline{6-8}
 & \NH & $\log \xi$ & $v_{\rm out}$ & & \NH & $\log \xi$ & $v_{\rm out}$ &$C$-stat/dof  \\ 
 &\scriptsize $10^{20}\ \rm cm^{-2}$ &\scriptsize  $\rm erg\ cm\ s^{-1}$ &\scriptsize  $\rm km\ s^{-1}$ & &\scriptsize $10^{22}\ \rm cm^{-2}$ &\scriptsize  $\rm erg\ cm\ s^{-1}$ &\scriptsize  $\rm km\ s^{-1}$ & \\
\hline
\hline 
 Sec 1 & $8\pm1 $   & $-0.9\pm0.2           $ & $-1400\pm200 $          & & $2^{+4}_{-1} $  & $3.9^{+0.2}_{-0.3} $    & $-76000^{+3000}_{-1000} $ & $849/684 $ \\
 Sec 2 & $11\pm1 $  & $-0.8_{+0.1}^{-0.2}   $ & $-1800\pm200 $          & & $8^{+8}_{-6} $  & $4.1\pm0.1  $           & $-76000\pm1000 $          & $772/689 $ \\
 Sec 3 & $10\pm1 $  & $-1.0\pm0.2           $ & $-1800\pm200 $          & & $3^{+7}_{-2} $  & $4.1^{+0.4}_{-0.2}   $  & $-75000^{+3000}_{-2000} $ & $800/686 $ \\
 Sec 4 & $9\pm1 $   & $-1.0\pm0.2   $         & $-2000\pm200 $          & & $4^{+3}_{-2} $  & $3.72^{+0.09}_{-0.07} $ & $-78300^{+500}_{-800} $   & $902/684 $ \\
\hline                  
\end{tabular}
\renewcommand{\arraystretch}{1.} 
\end{table*}

The EPIC-pn light curve of \zw shown in Figure \ref{fig:pn_lc} exhibits a double peak during the first observation \citep[see][]{Wilkins21} likely due to the intrinsic variability of the corona. This rapid change in flux represents a good probe to test if any outflow responds to the ionising luminosity on timescale of hours. Thus, we performed a time-resolved spectral analysis of the \zw observations taken in 2020. 

We split each observation into two sectors, obtaining in total 4 sectors of $\sim 40$ ks each. Both peaks are included in the second sector (see Section \ref{sec:obs_data}). In order to consider variation of the ionising luminosity in our spectral modelling of the outflow, we computed the SED for each sector following the same method presented in Section \ref{sec:sed}. We assumed that the UV flux does not change on this short time scale. \cite{Wilkins22} observed a variability at the 4 per cent level in the UV light curve of OM. Therefore, we simultaneously characterise the RGS and EPIC-pn spectra for each sub-epoch to study the short-term evolution of the ionised outflows of \zw.

To fit the spectra, we have taken advantage of the constraints we obtained from the fit of the time-averaged spectrum using the same continuum as a starting point. We kept fixed the parameters of the HIC to the values from Table \ref{tab:wa} because of its low significance and we investigated any evolution of the column density, ionisation parameter, and outflow velocity for the LIC and the UFO. 

In Figure \ref{fig:trs} we show the results of our time-resolved spectroscopy of the 2020 observation of \zw. We compare the evolution of the ionisation state, column density and outflow velocity of the LIC and UFO with the $0.3-10\,\rm{keV}$ light curve (top-panel). 

The ionisation parameter of the UFO (blue line in the second panel) drops between the second ($\log \xi=4.1\pm0.1$) and the last sector ($\log \xi=3.72\pm0.08$). Because of the large best-fit uncertainties obtained in the other sectors it is difficult to determine if there is a correlation or a time delay between the light curve and the ionisation parameter for the considered timescales. However, we noticed an increase of significance of the UFO component in the last sector, where the uncertainty of both the column density and the ionisation parameter is notably smaller. This is corroborated by the grid searches that we computed for each sector and displayed in Figure \ref{fig:contour} in the Appendix \ref{app:contour}. The $\NH$ and $\xi$ of the UFO in the fourth sector are better constrained and the their contour plot covers a smaller area in the parameter space. In this sector, the UFO component improves the overall fit by $\Delta C{\rm stat} = 50$.

Across the different time intervals, the outflow velocity of the UFO seems to follow an evolution similar to the ionisation state whereas the column density varies within the uncertainties. The LIC instead shows a column density evolution suggesting that a larger amount of material is found along the line of sight in correspondence with the highest flux. Regarding the ionisation state of the gas, the LIC does not vary within the errors whereas its outflow velocity increases between the first and second sectors. 

The best-fitting parameters for each sector are listed in Table \ref{tab:trs}. Overall the low signal-to-noise prevents an accurate time-resolved spectroscopy of the outflows. The present statistics together with the time gap between the two observations make it difficult to understand whether the outflows are in equilibrium with the ionising radiation or if the column density and outflow velocity vary on the considered timescales.

%--------------------------------------------------------------------
\section{Discussion}
\label{sec:discussion}
%--------------------------------------------------------------------

%-------------------------------------------------------------
%                                   Column density versus time 
%-------------------------------------------------------------
   \begin{figure}
   \centering
   \includegraphics[width=\hsize]{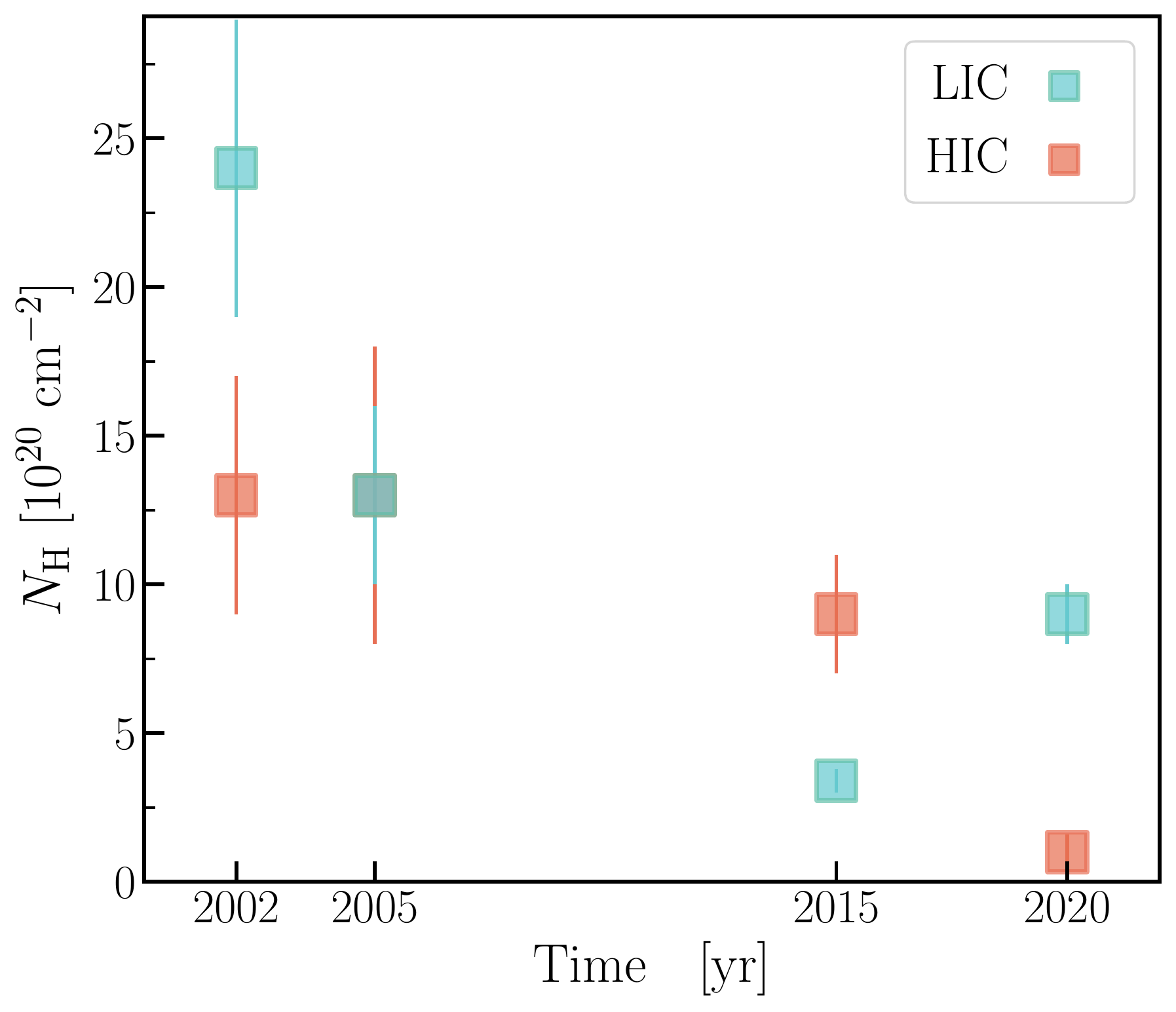}
      \caption{Hydrogen column density of the two warm absorbers through the years. The earlier values of the column density are taken from \citet{Costantini07} - 2002, 2005 - and \citet{Silva18} - 2015.
              }
         \label{fig:nh_t}
   \end{figure}

%-------------------------------------------------------------
%                              Outflow velocitites versus time 
%-------------------------------------------------------------
   \begin{figure}
   \centering
   \includegraphics[width=\hsize]{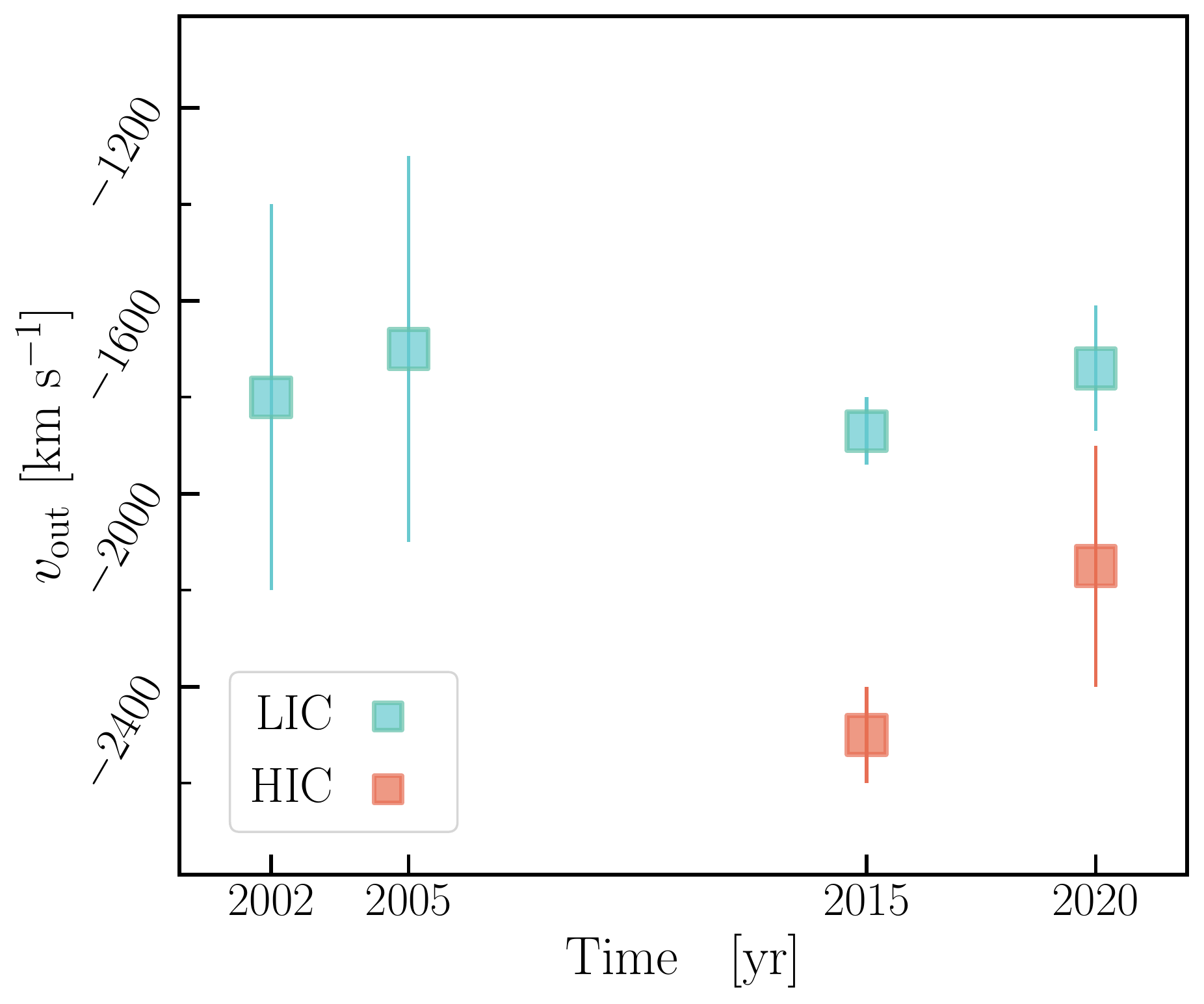}
      \caption{Long-term variability of the outflow velocity of the two warm absorbers. The velocities of the epoch $2002$ and $2005$ are taken from \citet{Costantini07} whereas the $2015$ velocities are from \citet{Silva18}. The outflow velocity of 2002 and 2005 are not plotted here since only upper limits were obtained in these two epochs. 
              }
         \label{fig:zv_t}
   \end{figure}

%-------------------------------------------------------------
%                             Luminosity versus ionisation par
%-------------------------------------------------------------
   \begin{figure*}
   \centering
   \includegraphics[width=.495\hsize]{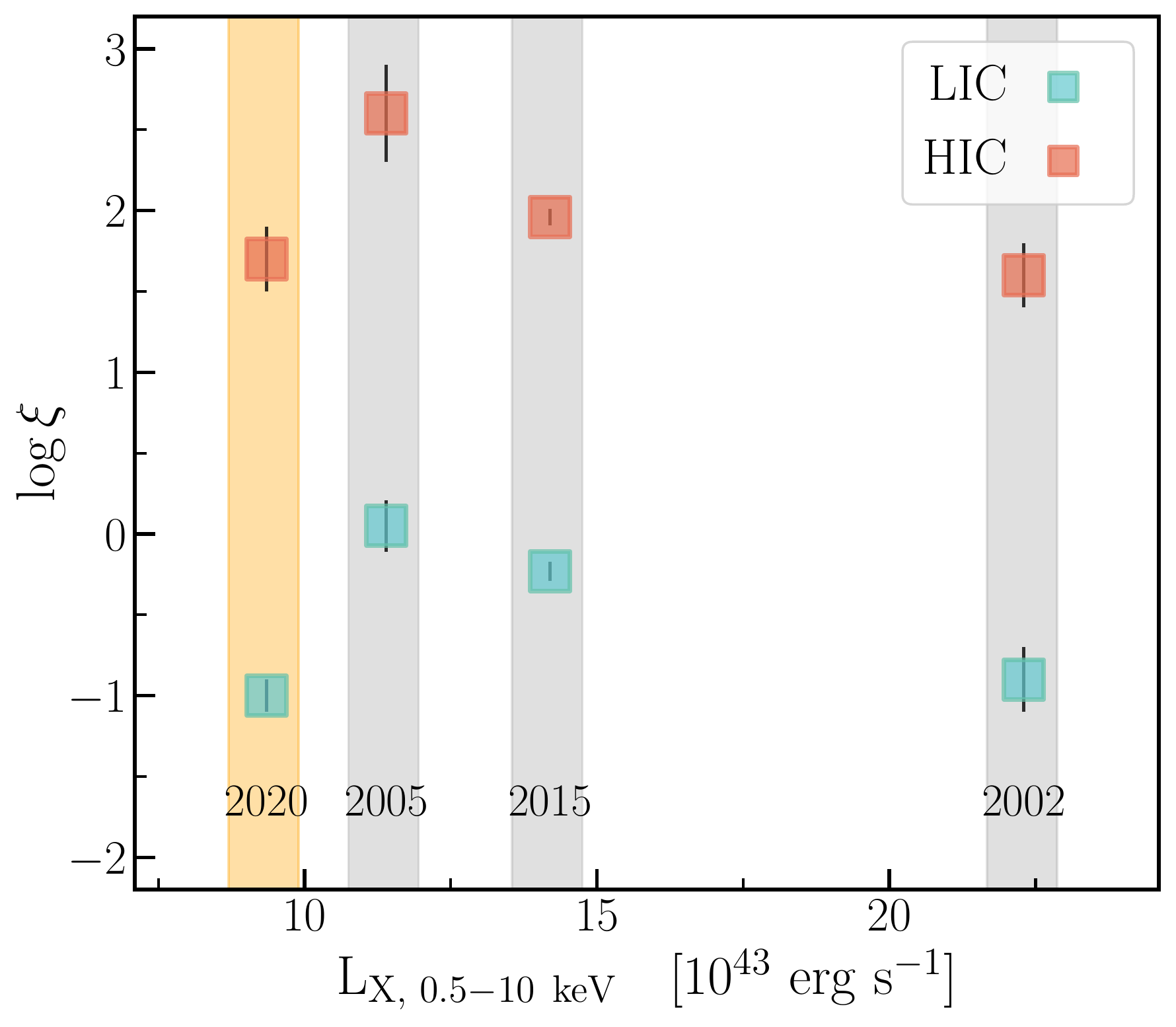}
   \includegraphics[width=.495\hsize]{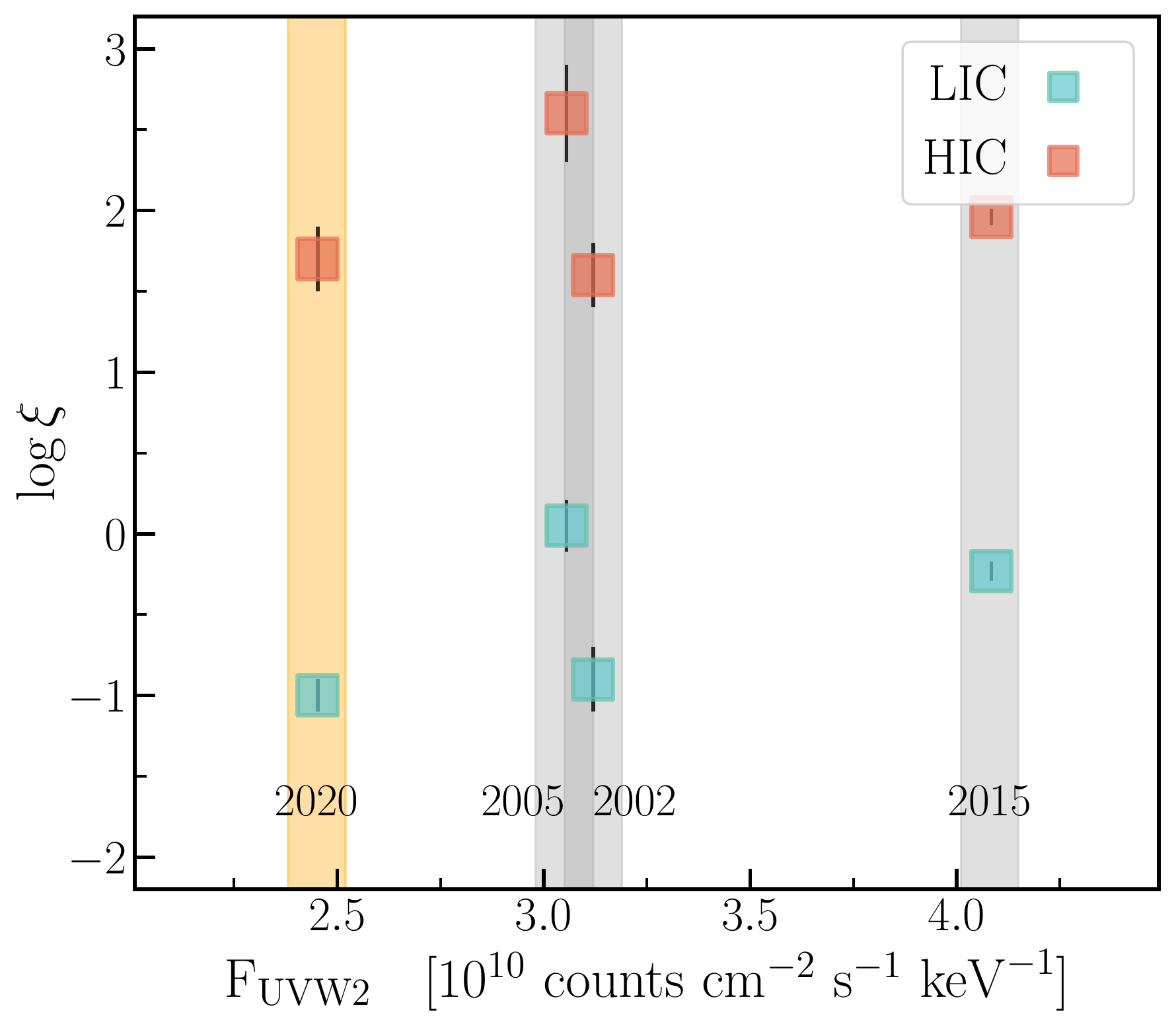}
      \caption{Long-term variability of the ionisation parameters as a function of the  X-ray luminosity (on the left) and UV emission (on the right). \textit{Left panel}: We updated figure 5 of \citet{Silva18} adding the last observation (at the time of writing) taken in 2020. Throughout the year the ionisation states of the outflowing plasma do not change accordingly with the X-ray ionising luminosity. \textit{Right panel}: for comparison purpose we plot the variability of the ionisation parameters as a function of UV emission observed with the UVW2 filter. We highlight the results of the present observation with the vertical yellow area whereas the results of previous epochs are indicated with the vertical grey bands. 
              }
         \label{fig:Lx_xi}
   \end{figure*}

%-------------------------------------------------------------
%                                                         SEDs 
%-------------------------------------------------------------
   \begin{figure}
   \centering
   \includegraphics[width=\hsize]{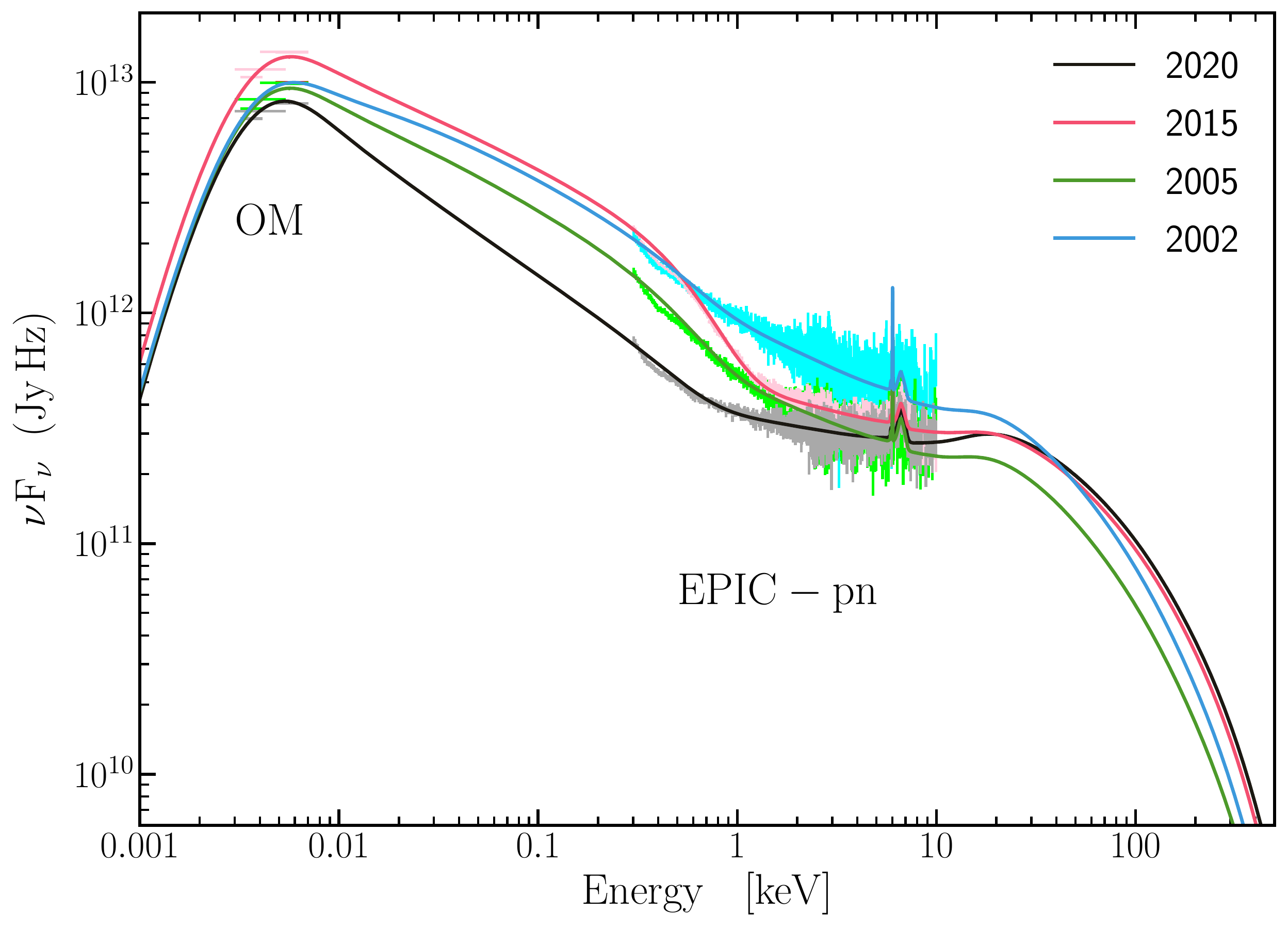}
      \caption{Comparison of the spectral energy distribution of \zw based on the OM and EPIC-pn data.
              }
         \label{fig:seds}
   \end{figure}

%-------------------------------------------------------------
%                              Density and thickness variation
%-------------------------------------------------------------
   \begin{figure*}
   \centering
   \includegraphics[width=.495\hsize]{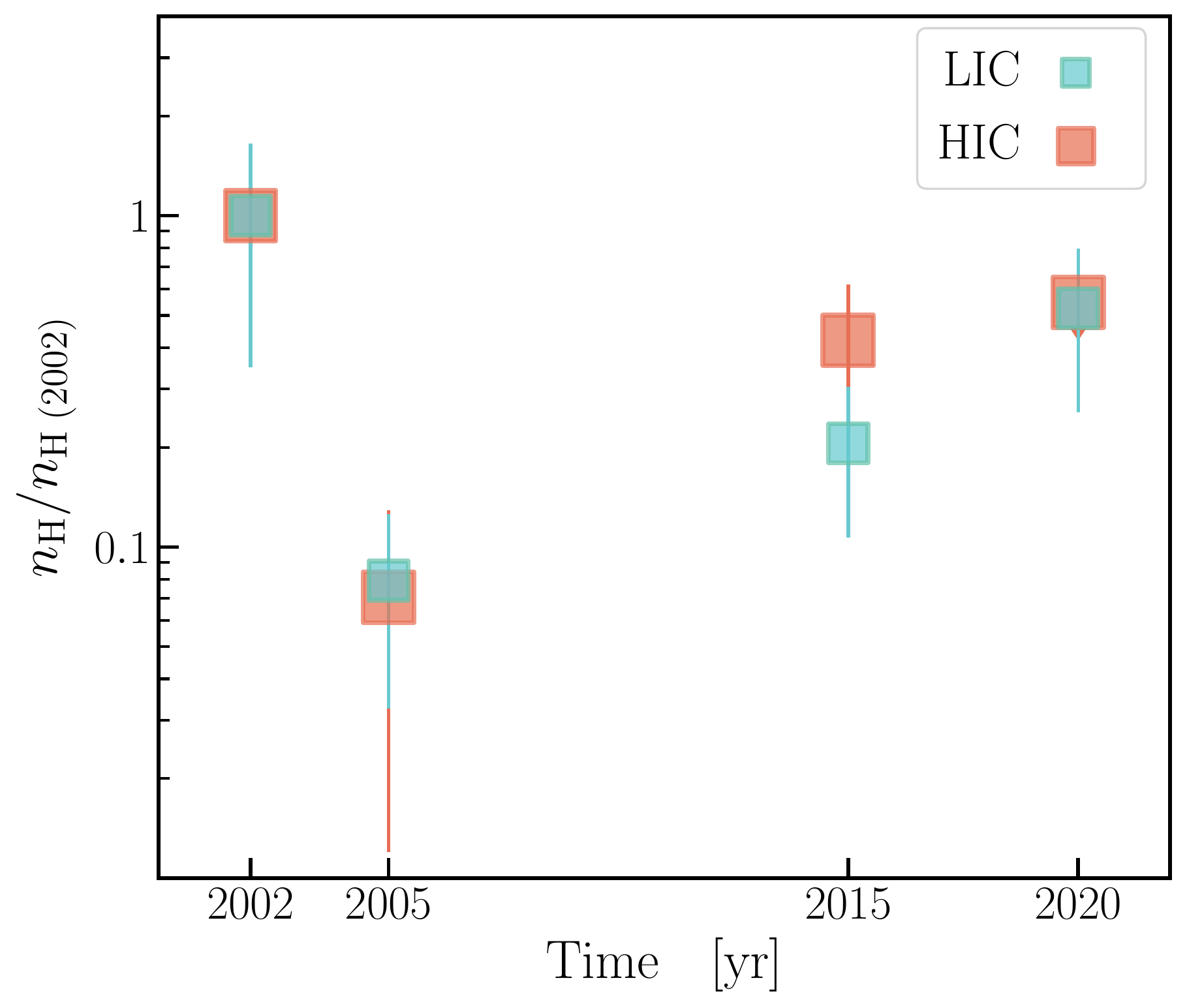}
   \includegraphics[width=.495\hsize]{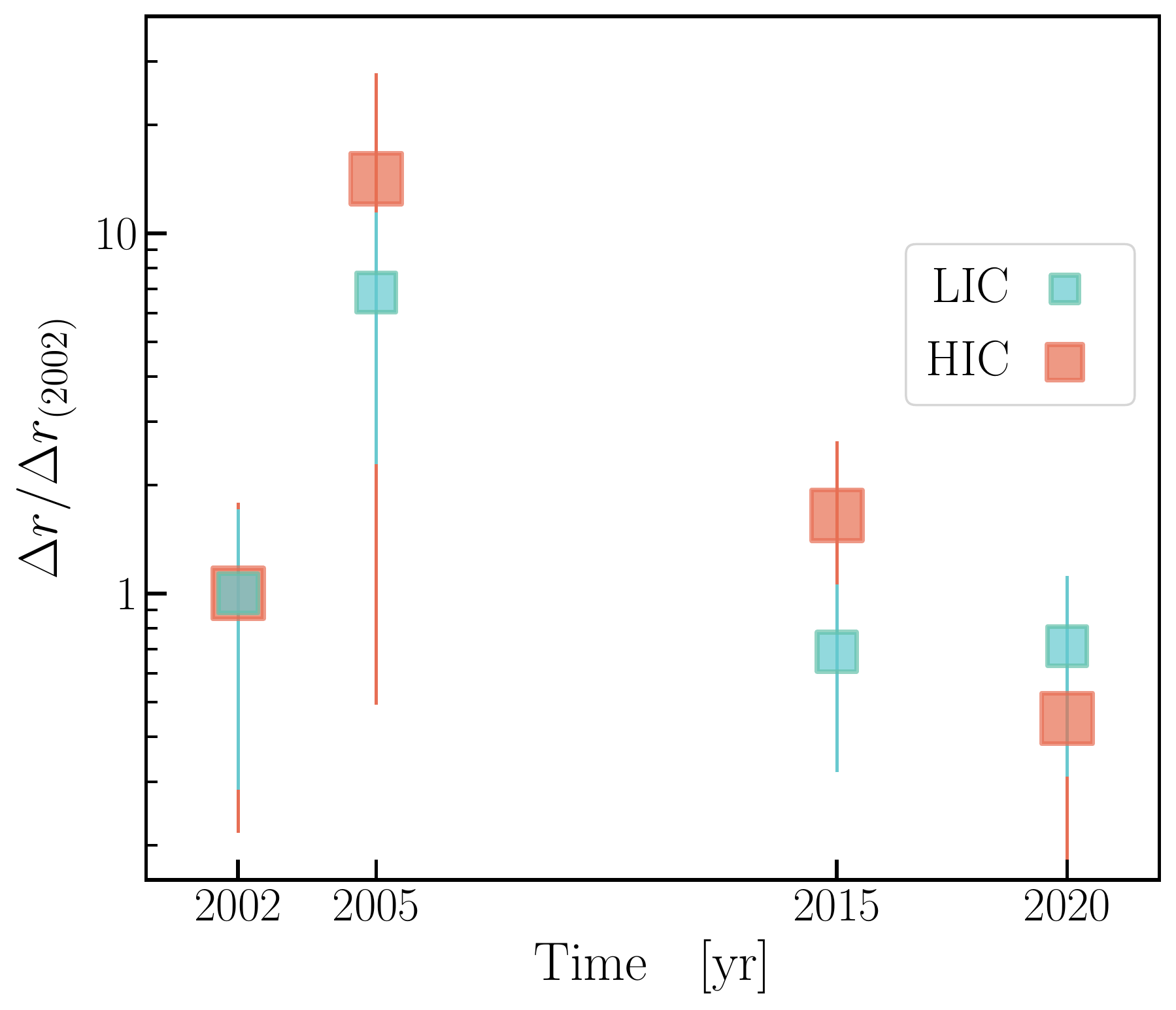}
      \caption{{\it Left panel}: Variation of volume density of the warm absorbers, $n_{\rm H}$, assuming the same distance. We normalised for the initial density. {\it Right panel}: Variation of the thickness, $\Delta r$ of the warm absorber through the different RGS epochs. We normalised for the observation taken in 2002. 
              }
         \label{fig:dn_dr}
   \end{figure*}

The recent RGS datasets taken in early 2020 show the presence of a two-zone warm absorber in the soft X-ray band of \zw. The results, presented in the previous sections, become particularly interesting when compared with the outflow properties observed in the previous epochs. Here, we discuss the long-term variability of the warm absorbers in \zw, their origin and energetics. In addition, the EPIC-pn spectra of the source endorse the presence of a persistent ultra-fast outflow already detected in early epochs. The light curve of the present datasets provides a test bench to investigate any short-term response of the gas to the ionising luminosity. In this section we will discuss our attempt to carry out a detailed time-resolved spectroscopy study.

\subsection{Variability of the warm absorbers}
During its 23 year long lifetime, \xmm observed \zw multiple times. The soft X-ray spectra retrieved in earlier epochs, 2002, 2005 and 2015, all required a two-component warm absorber, consisting of a low-ionisation and a high-ionisation plasma which are not in pressure equilibrium \citep{Costantini07,Silva18}. The 2020 dataset presented here shows the same phenomenology. 

In Figure \ref{fig:nh_t} we plot the variability of the hydrogen column density of the two-component warm absorber through the years. In the last observation the HIC (in red) shows a significantly low column density, which compromises its detection significance, with respect the previous epochs. The \NH of the LIC (in blue) varies substantially among the \xmm observations, dropping by at least a factor of five between 2002 and 2015 and increasing by a factor of three between 2015 and 2020. In contrast, the outflow velocity of both components have been constant over the past two decades (see Figure~\ref{fig:zv_t}). A similar velocity for the LIC was observed even at earlier epochs through UV spectroscopy \citep[$v_{\rm out} \sim1870\rm\ km\ s^{-1} $;][]{Laor97}. Due to the shorter exposure time, it has not been possible to constrain the flow velocity for the HIC in the first two epochs \citep{Costantini07}.

The ionisation state of both outflowing plasmas exhibits a peculiar long-term variability. In Figure~\ref{fig:Lx_xi} we plot the X-ray luminosity for all four epochs versus the ionisation state of the two warm absorbers. The X-ray luminosity alone represents a good diagnostic for the ionising luminosity since the ions in the X-ray band are more sensitive to high-energy photons \citep{Netzer03}. We also display how the ionisation state of the LIC and HIC relates with the UV flux observed by OM in the UVW2 filter, which is the only filter in common among all the observations. In both panels, the ionisation balance of two warm absorbers manifests either a complex or a stochastic time-dependent behaviour. Possible explanations to this non-standard evolution of the outflow ionisation states are given in Section \ref{sec:ori_wa}. 

The ionisation states of the warm absorbers detected in \zw clearly do not vary directly following the ionising luminosity. However, the ionisation state of the two components seems to vary in the same way through the different \xmm epochs. In the first three observations we observed an apparent anti-correlation between their ionisation parameters and the X-ray luminosity. In the 2020 observation, the LIC and HIC ionisation parameters and the X-ray luminosity drop close to their historical minimum in qualitative agreement with a standard scenario where a plasma in photoionisation equilibrium responds to the ionising flux.

Finally, based on the model proposed by \cite{Silva18} and summarised in the following section we can assume that the distance of the warm absorber does not change across the epochs. Thus, we can roughly estimate the absorber density variations using the definition of the ionisation parameter. For this reason it has been necessary to compute the ionising luminosity between 1-1000~Ryd for each epoch.

This quantity is not available for the previous observations. For each \xmm dataset, we uniformly characterised the accessible OM filters and the EPIC-pn spectrum using the SED model adopted in the present work. The best-fit SED from each epoch is compared in Figure \ref{fig:seds}. Both the shape and the normalisation of the SED of \zw clearly vary over year-long timescales. The densities of the LIC and HIC, shown in Figure~\ref{fig:dn_dr}, normalised for their value observed in 2002, seem to vary consistently on timescales of years. Starting from the density variation we can infer the changes in the warm absorber thickness, $\Delta r$, through the relation $\NH = f \cdot n \cdot\Delta r$ and assuming that the filling factor, $f$, remains constant over time. The right panel of Figure~\ref{fig:dn_dr} demonstrates that the thickness of the two warm absorber components seems to follow the same evolution within each other over the epochs.

The different ionisation measurements in the present work and the previous literature cannot be explained by a distinct SED modelling. \cite{Costantini07b} and \cite{Silva16} used indeed a pure phenomenological model fitting the UV and X-ray data by means of a piecewise powerlaw. Using the SEDs showed in Figure \ref{fig:seds}, we reanalysed the properties of the outflows detected in the older observations. There are no significant differences between our best values and the ones from the literature. The column density and ionisation parameter of LIC always vary within the uncertainties whereas we found slightly higher $\xi$ for the HIC (between 1 and 2.5$\sigma$). The different SED modellings have a negligible effect on our results about the long-term variability of the ionised absorbers.

We have also studied the short-term variability of the warm absorbers dividing the light curve in four sectors. Due to its low-significance, we fixed the HIC parameters and investigated the short-term behaviour of the LIC. If the LIC were in photoionisation equilibrium with the ionising luminosity we would expect to see the ionisation state of the LIC increasing by a factor of $\sim 2$ (0.3 dex) between the first and second sector following the change in flux. However, the large uncertainties on the inferred ionisation parameter values make it impossible to detect significantly if the gas responds or not to the ionising luminosity. Instead, both the column density and the outflow velocity of the LIC seem to change on time scale of hours. There is a tentative evidence that X-ray flares are accompanied by an increase of material and velocity. A similar short-term behaviour was observed by \cite{Silva18} for the HIC during the X-ray peak present in the observation taken in 2015. 

During the high-state observed in sector 2, the count rate is comparable with the count rate recorded in some segments of the 2015 observation \citep[in particular with the segment 2 in Table 3 of ][]{Silva18}. The ionisation parameters of LIC in sector 2 of 2020 ($\log \xi = -0.8_{+0.1}^{-0.2}$) and in segment 2 of 2015 ($\log \xi = -0.4\pm0.2$) appear consistent within $2\sigma$. However, both values are affected by large uncertainties which impede us from drawing any robust statement.

\subsection{Origin of the warm absorber in I Zw 1}
\label{sec:ori_wa}
The preliminary analysis of the HST/COS data of \zw taken during the multi-wavelength campaign of 2015 reveals the line-locking of the \nv $\lambda1239,1243$ doublet lines \citep{Silva18}. This warm absorber observed in the UV is believed to be the counterpart of the LIC in the X-rays \citep{Costantini07}. The locked \nv lines would provide important hints about the acceleration mechanism of the LIC. Line-locking systems are usually believed to be a signature of outflows driven by radiation forces. Lines formed in different outflow clumps can become locked at the doublet line separation due to shadowing effects in radiatively driven outflows. Under certain circumnstances, clouds closer to the origin of the outflows can shield clouds further out resulting in a synchronising of the outflow velocity \citep{Milne26,Scargle73,Braun89}. 
Whereas the \civ line-locking feature is largely visible in both broad absorption quasars (BALs) and non-BALs quasars, the line-locking feature from \nv is often non detected \citep{Bowler14,MasRibas19}. Therefore, the outflows of \zw might represent a peculiar system both in the UV and in the X-rays. The UV data will be presented in detail by our team in a future paper. 

To explain the non-trivial behaviour of the ionisation states of the absorbers as a function of luminosity, \cite{Silva18} proposed a phenomenological model in which different clumps, with different density, cross the observer's line of sight at different epochs. In this configuration, the ionisation state is driven by the different densities of the clouds and not by the ionising luminosity \citep{Dyda18}. In this respect, a clumpy outflow can naturally explain the changes in opacity of the warm absorbers observed in the different datasets. 

\cite{Silva18} suggested that the two-phase warm absorber arises from the same clump, with the HIC facing the central source, since their ionisation parameters followed the same evolution. Our results from the time-averaged analysis of the 2020 \xmm campaign support the clumpy configuration suggested by \cite{Silva18}. The fact that the outflow velocity of both warm absorber components did not change indicates that we are staring at the same ejection phenomenon through throughout all the epochs. The slightly higher velocity of the HIC agrees with the standard radiatively-driven wind model \citep{Proga98,Waters21} where the less-dense, higher velocity, stream is confined at the more distant side of a denser, colder and slower outflow. The connection between the two components would also provide a natural explanation for the similar evolution of the density and thickness of the two components over the last twenty years (Figure \ref{fig:dn_dr}).

As an alternative, the odd behaviour of the warm absorber in \zw could be explained by a scenario in which the outflowing absorbing gas is in persistent non-equilibrium with the ionising source. In general, for a gas in photoionisation equilibrium, the plasma responds instantaneously to a variable source luminosity becoming more ionised as the flux increases and recombining when the flux of the source drops. In the presence of a low-density gas, its response to the change in luminosity may either be delayed or negligible \citep[][]{Nicastro99,Kaastra12,Silva16}. Unfortunately, the history of the ionising luminosity before each considered epoch is not known. No \rxte, \swift, \nicer monitoring has ever been planned before the \xmm epochs. Therefore, it is not possible to infer the typical time delay of the warm absorbers with a time-dependent photoionisation modelling (Rogantini et al. submitted).

\subsection{Ultra-fast outflow detection}
The coexistence of warm absorbers and ultra fast outflows has been observed along the line of sight of multiple AGN. \cite{Tombesi13} noted that ~70\% of the Seyfert 1 galaxies in their sample with a detected ultra-fast outflow also show a warm absorber. \zw can be added to this subset of sources since already earlier \xmm/EPIC-pn observations reveal the presence of such a fast wind \citep{Reeves19}. In the observation of 2015, the presence of a blueshifted absorption line at 9~keV suggests the presence of an accretion disc wind with an outflow velocity of $\sim 0.26c$ and ionisation parameter $\log \xi = 4.9$. Analysing the broad and P-Cygni iron K profile of the 2002, 2005, and 2015 observations, \cite{Reeves19} did not observe any significant long-term variability in those datasets though wind parameters estimated in the earlier epochs are affected by large uncertainties.

In our fit, the outflow velocity of the gas is unchanged with respect to previous epochs ($v_{\rm out} = -77100\pm 400\ \rm km/s$; $\sim0.26c$), while its ionisation state is significantly lower ($\log \xi = 3.80_{-0.04}^{0.11}$). The best-fit value of the ionisation parameter is driven by the shape and the absorption features of iron, silicon, and sulfur at lower energies (in Figure \ref{fig:wa} and \ref{fig:wa_pn}). In the Fe K band we did not detect any statistically significant absorption feature at 9~keV. Instead, the best-fit model highlights the presence of an absorption line at a lower energy of 8~keV.

Similarly to the warm absorber, the outflow velocity of the UFO does not show long-term variability. We are likely observing at the same wind which has changed its ionisation state between 2015 and 2020. Between the two epochs the ionising luminosity dropped by a factor of $\sim 2$; thus, it cannot fully motivate the one-order of magnitude decrease observed for the ionisation parameter. A clumpy system where the ionisation state is density driven could explain the large variation of the ionisation parameter. In the 2015 observation, \cite{Reeves19} found a discrepancy between the ionising luminosity predicted by their radiative disc wind modelling and versus the observed luminosity. An inhomogeneous and clumpy wind represents one of the explanations proposed by the authors to explain the discrepancy without requiring the intrinsic X-ray luminosity to be suppressed. Finally the probability distribution of the wind outflow velocity computed in the Bayesian data analysis shows multiple peaks (see Figure \ref{fig:bay}). This also might suggest the presence of a system with a complex outflow velocity distribution rather than a single speed wind. The low signal-to-noise ratio prevents us from investigating it further.  

A short-term variability of the ionisation state of the UFO is also observed. During the X-ray flares the wind shows a higher ionisation state with respect the second dataset (in particular sector 4) which comes after the flaring period. These results are suggestive that the ionisation state of the UFO responds to the flaring in the X-ray emission on short timescales. \cite{Wilkins22} also noticed a variable highly ionised absorption line in the Fe K band for the same datasets using a Gaussian line scanning \citep[see][]{Pinto16}. In the 2015 observation, the opacity of the wind appears to be anti-corralated with the X-ray flux on short timescales \citep{Reeves19}. Here, the opacity change could either be due to a response in ionisation of the wind, as we might observe in the present observation, or via modest variation of the wind column density. 

\subsection{Outflow energetics in I Zw 1}
In contrast to other sources, \zw has proved to host a peculiar warm absorber system, whose geometry may differ from a classical conical structure where the AGN outflows are stratified according to ionisation and distance. Therefore, directly connecting the UFO and and the warm absorber as layers of the same structure \citep[e.g][]{Tombesi13} may not be appropriate for this source.

The distance of the absorber from the central source is a crucial parameter to estimate the mass and the kinetic energy carried via the outflow per unit time. Since it is impossible to obtain the line-of-sight projected location of the absorbers through X-ray spectroscopy, unless time-dependent effects are included in the analysis \citep[e.g.][]{Nicastro99}, we constrained it using the assumptions and formulas presented in \cite{Blustin05}. The lower limit on the launching radius, $r_{\rm min}$, can be estimated assuming that the outflow has to obtain a speed greater than or equal to the escape velocity:
\begin{equation}
r_{\rm min} \ge 2GM_{\rm BH}/v_{\rm out}^2 
\end{equation}
where $G$ is the gravitational constant, $M_{\rm BH}$ is the mass of the black hole in \zw which is estimated to be between $8\times 10^6$ and $3\times10^7\,M_{\odot}$ using the $\rm H_{\beta}$ line and optical reverberation mapping \citep[][]{Vestergaard06,Huang19}. By substituting the outflow velocity of $v_{\rm out}=0.26c$ observed both in the 2020 observation and in the previous epochs, a lower limit of $r_{\rm min} = 4\times 10^{-5}\; \rm pc\simeq 14\ \emph{r}_{g}$ is found for the UFO \citep[assuming a black hole mass of $3\times10^7\,M_{\odot}$;][]{Wilkins21,Wilkins22}. Instead adopting the outflow velocity observed for the warm absorbers we obtain a higher launching radius, $r_{\rm min} \sim0.2\: \rm pc$, for the slower wind.

We computed the energy budget of the UFO following the same approach presented in \cite{Reeves19}. In particular, we derived a conservative estimate of the mass outflow rate $\dot M > 0.004 \rm \; M_{\odot} \rm \; yr^{-1}$ for the smallest inner radius of the wind. Normalising this quantity to the Eddington rate we obtain $\dot M > 0.006 \; \dot M_{\rm Edd}$ which is one order of magnitude lower than the mass outflow rate observed in 2015 \citep{Reeves19}. This also leads to a smaller value of kinetic power, $\dot E_{\rm k} > 6\times10^{42}\; \rm erg\; s^{-1}$ ($\dot E_{\rm k} > 0.002 \: \dot  L_{\rm Edd}$) with respect to the previous epoch.

In 2010, \cite{Cicone14} observed \zw with the submillimeter IRAM telescope to investigate the presence of molecular gas on kiloparsec scales. The CO observations showed no clear evidence of any large-scale molecular outflow. Subsequently, the authors determined an upper limit of $140\: \rm  M_{\odot}\: yr^{-1}$ for the total molecular mass outflow rate which corresponds to an upper limit on the kinetic power of $\dot E_{\rm k} < 10^{43} \; \rm erg\; s^{-1}$ ($\dot E_{\rm k} < 3\times10^{-3}\;  L_{\rm Edd}$ in Eddington units) assuming a conservative value for the outflow velocity, $v_{\rm out}<500\; \rm  km\; s^{-1}$. This constraint on the molecular outflow luminosity is more than an order of magnitude lower than the kinetic power of the ultrafast X-ray wind, $\dot E_{\rm k} = (2-5)\times 10^{44}\ \; \rm erg\; s^{-1}$, observed in 2015 by \cite{Reeves19}. They concluded that at larger scales an energy-conserving molecular outflow could be ruled out suggesting, thus, a low efficiency in transferring the kinetic energy of the inner wind out to the kiloparsec molecular component. 

However, the 20-fold decrease of the UFO kinetic power observed in the \xmm observation of 2020 complicates the outflow scenario in \zw. We found that the conservative value on the kinetic luminosity of the fast X-ray wind is in agreement with the upper bound on the molecular outflow. Therefore, such strong variability of the ultrafast outflow prevents us from ruling out the presence of an energy-conserving wind in \zw. 

\section{Summary}
\label{sec:conclusion}
In this investigation we studied the latest \xmm observation of the narrow-line Seyfert 1 galaxy \zw in which the X-ray flux has been found at its historical minimum. In particular we determined the SED of the source by taking into account all the non-intrinsic UV-X-ray processes along the line of sight; characterised the AGN outflows through photoionisation modelling and \xmm RGS spectroscopy; studied the long- and short-term variability of the outflows; and estimated their energetics. From our extensive modelling of the spectral features through spectral fitting we conclude the following:
\begin{description}
\item -- The AGN outflow in \zw consists of a two-component warm absorber and an ultra-fast outflow as previously observed by \cite{Costantini07}, \cite{Silva18} and \cite{Reeves19} in earlier observations of \zw. The first warm absorber component exhibits a low ionisation state, $\log \xi = -1.0\pm 0.1$, and an outflow velocity of $v_{\rm out}=-1750\pm 100 \rm \ km \ s^{-1}$. A higher ionisation state, $\log \xi = 1.7\pm 0.2$, and larger flow velocity, $v_{\rm out}=-2150^{+200}_{-250} \rm \ km \ s^{-1}$ characterise the second component. This component has a lower opacity and it is statistically less significant than the LIC. The UFO shows a high ionisation state, $\log \xi = -3.80_{-0.04}^{+0.11}$ and an outflow velocity of $v_{\rm out}\sim0.26c$.
\item -- On timescales of years, the ionisation states of both LIC and HIC do not correlate with the ionising luminosity, disagreeing with a photoionisation equilibrium scenario. The column density of the LIC also varies among different epochs whereas its outflow velocity remains unchanged since its first detection in the UV \citep{Laor97}. Similarly, the outflow velocities of the HIC and the UFO did not show any significant variation during the last two observations while the ionisation state of the UFO dropped by a factor of 10. 
\item -- In order to explain the odd response of the warm absorbers to the variability of the ionising source, the data support the scenario proposed by \cite{Silva16} in which the observed changes in ionisation are driven by the density of the gas. Clouds with different densities, hence different ionisation states, belonging to a persistent inhomogeneous clumpy outflow can justify the odd long-term variability of the warm absorbers in \zw. The difference in their ionisation state and their similar outflow velocity would suggest that the warm absorbers arise in two different regions of the same clump, with the lower ionisation component located to the far-away side of it and shielded by the higher ionised absorber. The line-locked \nv system observed in the UV  indicates that the radiation pressure might drive the LIC whose opacity is high enough \citep{Proga04}. Alternatively, we may be observing a warm absorber which is always out of ionisation equilibrium with the ionising continuum.
\item -- We calculated the energetics of the ultra-fast outflow and compared them with the previous epochs. Both the outflow mass rate and the kinetic power of the wind decreased by a factor of $\sim 20$ in 2020. This strong long-term variability of the UFO prevents us from drawing any conclusion regarding the presence of a large-scale, energy conserving wind in \zw. 
\end{description}

A dedicated monitoring campaign of the ionising luminosity followed by a long \xmm observation of \zw would help to understand the nature of its outflows and why their ionisation parameters do not correlate with the ionising continuum. Moreover, the large effective area of future high-resolution X-ray spectrometers, such as \xrism/Resolve \citep{Ishisaki18}, \athena/XIFU \citep{Barret16} and \arcus \citep{Smith16} will enable a time-resolved spectroscopy analysis over short timescales. 

\section*{Acknowledgements}
We thanks the anonymous referee for their valuable feedback on the original version of this manuscript. We would like to thank Erin Kara, and Peter Kosec for many useful discussions. Support for this work was provided by NASA through the Smithsonian Astrophysical Observatory (SAO) contract SV3-73016 to MIT for Support of the Chandra X-Ray Center (CXC) and Science Instruments. The CXC is operated by the Smithsonian Astrophysical Observatory for and on behalf of NASA under contract NAS8-03060. DRW acknowledges support from the NASA NuSTAR and XMM-Newton Guest Observer programs under grants 80NSSC20K0041 and 80NSSC20K0838. WNB acknowledges support from NASA grant 80NSSC20K0795 and the V.M. Willaman Endowment. This research has made use of the NASA/IPAC Extragalactic Database, which is funded by the National Aeronautics and Space Administration and operated by the California Institute of Technology. Foundation. This work has made use of the SIMBAD database, operated at CDS, Strasbourg, France and it is based on observations obtained with XMM-Newton, an ESA science mission with instruments and contributions directly funded by ESA Member States and NASA.

\section*{Data Availability}
The dataset studied in this work are available in the \xmm and \nustar public archives. The \xmm observations can be granted via the XMM-Newton Science Archive \url{http://nxsa.esac.esa.int/nxsa-web} searching for obsIDs 0851990101 and 0851990201. The \nustar dataset can be accessed using the NASA HEASARC Data Archive webpage (\url{https://heasarc.gsfc.nasa.gov/docs/archive.html} via obsID 60501030002. The X-ray spectra were analysed using the fitting program \spex (\url{https://doi.org/10.5281/zenodo.4384188}) and the Bayesian inference tool \textsc{MultiNest} (\url{https://github.com/JohannesBuchner/MultiNest}).



\bibliographystyle{mnras}
\bibliography{/Users/danieler/ownCloud2/paper_latex/bibliography/biblio} % if your bibtex file is called example.bib

\appendix
\section{Bayesian data analysis}
\begin{figure*}[]
   \centering
   \includegraphics[width=\hsize]{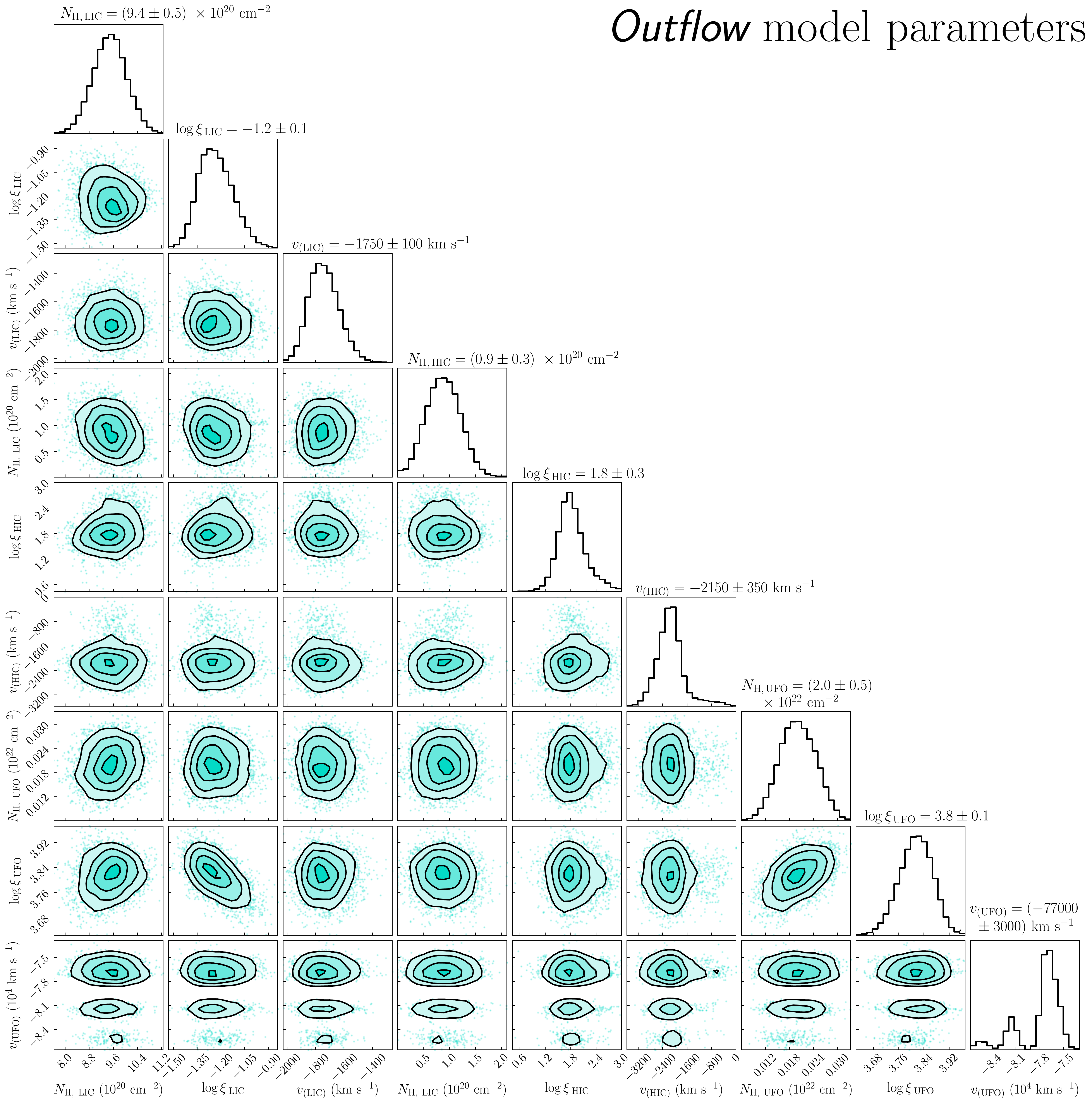}
      \caption{One- and two-dimensional marginal posterior distributions of the outflow parameters. In order from top to bottom (and from left to right) the ionisation parameter, column density and outflow velocity of the LIC, HIC and UFO. In the two-dimensional histograms the contours indicate the 1$\sigma$, 2$\sigma$, 3$\sigma$, 4$\sigma$ confidence intervals.
              }
         \label{fig:bay}
   \end{figure*}

To fully explore the complex parameter space of the different outflows we used a Bayesian data analysis. The nested sampling algorithm \textsc{MultiNest} has been used to infer the parameters of the photoionisation models. In Figure \ref{fig:bay} we display the distributions obtained using three photoionisation models which confirm the presence of the two-component warm absorber and the ultra fast outflow. We used uninformative priors and we fixed the turbulence velocity to the default value of $100\ \rm km\ s^{-1}$.

%-------------------------------------------------------------
%                             Luminosity versus ionisation par
%-------------------------------------------------------------

\section{Time-resolved analysis of the UFO}
\label{app:contour}

%-------------------------------------------------------------
%                             Luminosity versus ionisation par
%-------------------------------------------------------------
   \begin{figure*}[]
   \centering
   \includegraphics[width=\hsize]{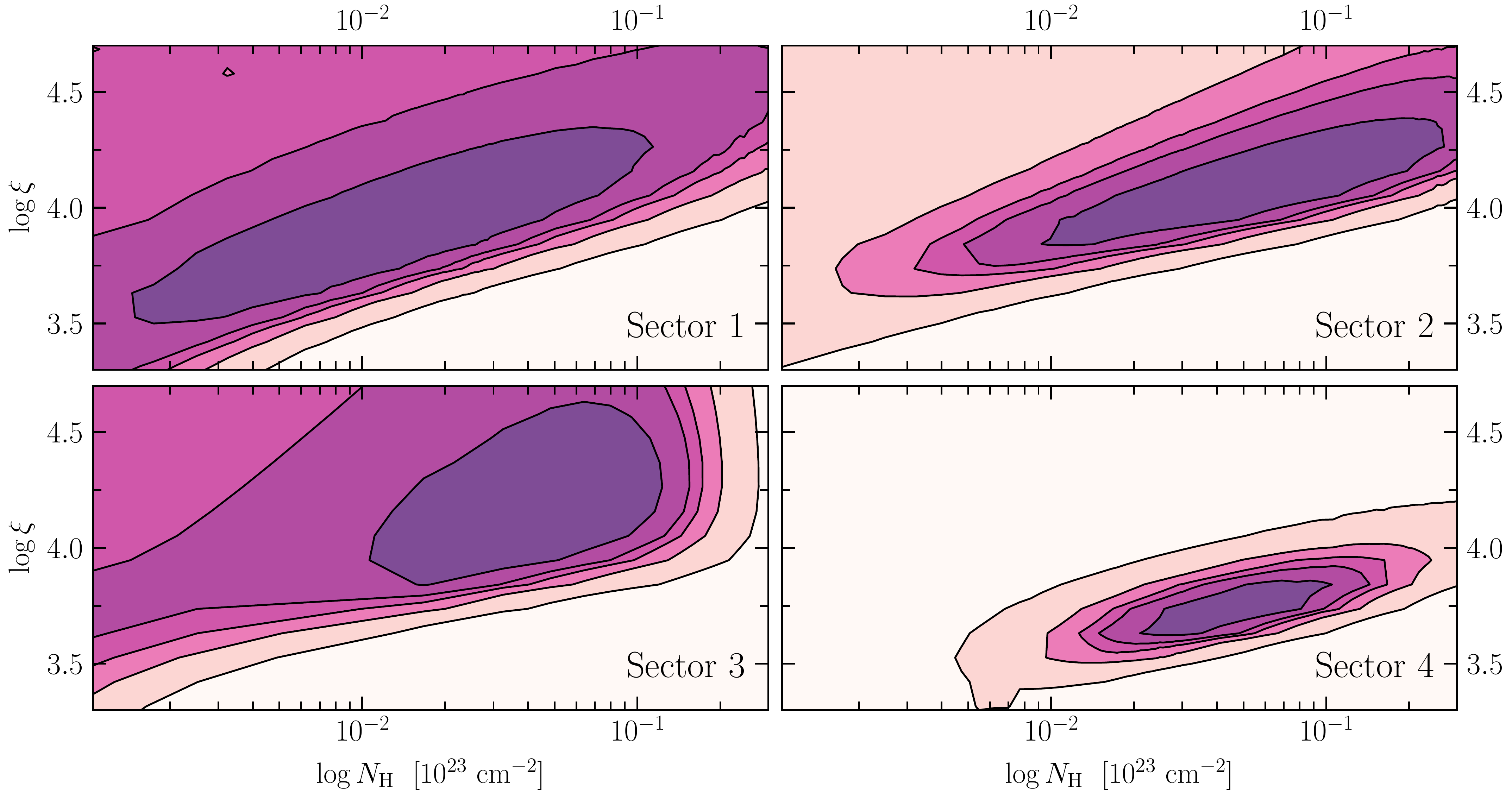}
      \caption{Search grid of the UFO \NH and $\log \xi$ performed for each sector of the time-resolved analysis. The contours correspond to 68.3\%, 90\%, 95.4\%, 99\% and 99.99\% confidence levels.
              }
         \label{fig:contour}
   \end{figure*}

The significance of the detection of the UFO in the average spectrum of the present \xmm observation is relative low. In time-resolved analysis, the search for the best values of the UFO parameters and their uncertainties is complex. Therefore, we performed a two dimensional grid search for the column density and ionisation parameter of the UFO in each sector. In this way we can also check if there is a complicate correlation between the two parameters. We considered the interval $(10^{-3}-0.1)\ \rm cm^{-2}$ for the column density and $3-5$ for $\log \xi$. In Figure \ref{fig:contour}, we show the plot contour for each sector.

% Don't change these lines
\bsp  % typesetting comment
\label{lastpage}

\end{document}